\documentclass[12pt,letterpaper]{JHEP3}  
\pdfoutput=1
\usepackage{amssymb,amsfonts}

\usepackage{epsfig}

\def\be{\begin{eqnarray}}
\def\ee{\end{eqnarray}}
\newcommand{\nn}{\nonumber}
\newcommand\para{\paragraph{}}
\newcommand{\ft}[2]{{\rm{\frac{#1}{#2}}}}
\newcommand{\eqn}[1]{(\ref{#1})}

\def\Dslash{\,\,{\raise.15ex\hbox{/}\mkern-12mu D}}
\def\Dbarslash{\,\,{\raise.15ex\hbox{/}\mkern-12mu {\bar D}}}
\def\delslash{\,\,{\raise.15ex\hbox{/}\mkern-9mu \partial}}
\def\delbarslash{\,\,{\raise.15ex\hbox{/}\mkern-9mu {\bar\partial}}}
\def\pslash{\,\,{\raise.15ex\hbox{/}\mkern-9mu p}}
\def\calDslash{\,\,{\raise.15ex\hbox{/}\mkern-12mu {\cal D}}}

\def\lae{\mathrel{\mathop{\smash{\lower .5 ex \hbox{$\stackrel<\sim$}}}}}
\def\lae{\mathrel{\mathop{\smash{\lower .5 ex \hbox{$\stackrel>\sim$}}}}}

\title{Gauge Dynamics and Topological Insulators}

\author{Benjamin B\'{e}ri${}^1$, David Tong${}^2$ and Kenny Wong${}^2$\\

${}^1$ TCM Group, Cavendish Laboratory, University of Cambridge, UK\\
${}^2$ Department of Applied Mathematics and Theoretical Physics, University of Cambridge, UK \\ 

{\tt bfb26@cam.ac.uk, d.tong@damtp.cam.ac.uk, k.wong@damtp.cam.ac.uk}
}

\abstract{A non-abelian magnetic field in Yang-Mills theory induces the formation of a ``W-boson'' vortex lattice. We study the propagation of fundamental fermions in the presence of this lattice in $2+1$ dimensions.  We show that the spectrum for massless fermions contains four topologically-protected Dirac points with non-zero Bloch momentum. For massive fermions, we compute topological invariants of the band structure and show that it is possible to realise a $\mathbb{Z}_2$ topological insulator within \mbox{Yang-Mills theory.}}

\begin{document}
\pagestyle{plain} \setcounter{page}{1}
\newcounter{bean}
\baselineskip16pt

\newpage

\section{Introduction}

It was discovered long ago that the presence of a background magnetic field in a non-abelian gauge theory results in a tachyonic mode for the gauge bosons \cite{no}. Upon condensation, they form a vortex lattice.  Many properties of this instability and the resulting ground state have been studied, most notably  in a series of papers by Ambj\o rn and Olesen \cite{ao0}-\cite{ao4}.

\para
We know from condensed matter physics that the presence of a lattice  gives rise to a range of beautiful phenomena, from elementary band structure to the emergence of novel topologically protected quantities. The purpose of the present paper is simply to understand some of these properties in Yang-Mills  theory by studying the propagation of fundamental fermions in the background of the magnetically-induced gauge boson lattice. We will see how a number of aspects of topological insulators naturally arise in non-Abelian gauge theories. 

\para
In pursuing this problem, there is one immediate obstacle: the full solution for the background lattice is known only numerically. Fortunately, the  topological nature of our quest comes to the rescue. We will show that a number of the key features of the fermion dynamics can be computed exactly, even though one doesn't have full control over the background in which they move. These properties, which are protected by a combination of topology and symmetry, will be the focus of this work.

\para
While much of the work on topological insulators focusses on theories with a gap, there are also interesting things to say when the gap closes. In our case, we will show that, in the absence of anomalies, Dirac points in the W-boson lattice necessarily come in multiples of four.  Perhaps somewhat surprisingly, these Dirac points do not lie at zero Bloch momenta $\vec{p}$. Instead, they are shifted to $|\vec{p}|\sim \sqrt{B}$, where $B$ is the magnitude of the magnetic field. We will show that the  position of these Dirac points is one of properties that can be computed exactly: assuming certain symmetries, their locations are immune to both deformations of the lattice and quantum corrections. 

\para
For massive fermions propagating in the background of the lattice, the band structure has a gap. We are now firmly in the realm of topological insulators, 
where a number of topological invariants are associated to the  band structure of gapped systems. (See, for example, \cite{hk,zreview} for reviews). The presence of a background lattice emerging naturally in Yang-Mills theory provides an opportunity to explore these ideas in a context familiar to high-energy theorists. 
To this end, we provide explicit calculations of the first Chern numbers, known as TKNN invariants \cite{tknn}, as well as the 
more recently discovered $\mathbb{Z}_2$ invariant \cite{km} of the lowest bands. We will see that the resulting phenomenology of this phase is essentially that of quantum spin-Hall physics. Moreover, we will show that $SU(2)$ Yang-Mills coupled to fundamental fermions contains within it the simplest $\mathbb{Z}_2$ topological insulator\footnote{It has recently been argued that QCD in $d=3+1$ dimensions also lies in a topological phase \cite{ariel}.}.

\para
We start in Section \ref{sec2} by reviewing the gauge boson instability. We present a linearised approximation to the full lattice solution. As we explain, this linearised approximation is not a particularly good approximation. Nonetheless, it will prove sufficient for our needs, acting as a springboard for the exact results that follow.

\para 
Section \ref{fermionsec} studies the dynamics of  massless fermions in this lattice. We diagonalise the fermions in the background of the lattice and compute the location of the Dirac points, before showing that the existence and location of these points is exact. 
We also find something rather cute in the linearised regime:   the off-diagonal gauge boson field --- which we refer to as the ``W-boson"   $W(\vec{x})$ ---  plays a dual role. It not only describes the spatial lattice, but also, with a suitable rescaling, 
determines the energy over the Brillouin zone, $E=W(\vec{p})$.

\para
Section \ref{topsec} deals with massive fermions. We explain in detail how to quantize the fermions, pay particular attention to the discrete symmetries of the theory and provide detailed calculations of the topological invariants. We also pause in number of places to offer a translation between high-energy and condensed matter language. In particular, we will see how one can vary the chemical potential of the system, to move between trivial and non-trivial $\mathbb{Z}_2$ topological insulators.  Finally, a number of more involved calculations are relegated to a series of appendices.

\section{Non-Abelian Magnetic Fields}\label{sec2}

Throughout this paper we discuss $d=2+1$ dimensional $SU(2)$ Yang-Mills coupled to a background Higgs field and Dirac fermions\footnote{Our conventions: We use signature $(-,+,+)$. The  non-Abelian field strength is $F_{\mu\nu} = F_{\mu\nu}^aT^a$ where the $SU(2)$ generators are $T^a=\frac{1}{2} \sigma^a$. When we introduce fermions in Section \ref{fermionsec}, we use the representation of gamma matrices $\gamma^\mu = \{i\sigma^3, \sigma^1,\sigma^2\}$.}. The fermions are our primary interest and will be discussed in Sections \ref{fermionsec} and \ref{topsec}. The purpose of this section is to introduce the bosonic background in which the fermions propagate. This is a solution to the action
\be
S_{\rm bosonic} = - \int d^3x\  \ \frac{1}{2e^2}  {\rm Tr}\, F_{\mu\nu} F^{\mu\nu}+ \ {\rm Tr}\, D_\mu \phi D^\mu \phi  + \frac{g}{4} \left({\rm Tr}\,\phi^2 - \frac{v^2}{2}\right)^2  \label{action} \ee
where $\phi=\phi^aT^a$ is a real, adjoint-valued scalar field. 

\para
We will restrict ourselves to the regime  $v\gg e $ where the theory is weakly coupled, and we assume that $g \gg e^2$ where fluctuations of the Higgs field are suppressed. The $SU(2)$ gauge group is broken to $U(1)$ and the low-energy excitations consist of a photon and charged W-bosons of mass $m_W=ev$.  We decompose the gauge field as
\be A_\mu = \frac{1}{2}\left(\begin{array}{cc} a_\mu & W_\mu\nn\\ W^\star_\mu & -a_\mu\end{array}\right)\nn\ee
Throughout this paper, we are interested in what becomes of excitations in the presence of a background magnetic field,
\be F_{12} = \frac{B}{2}\sigma^3\label{nonabmag}\ee
Before we get into details, it's useful to first build  a simple, intuitive picture for what's going on. In a magnetic field all charged states form Landau levels.  Consider a scalar field of charge $q$ and mass $m$ in a magnetic background $B$. The familiar Landau level quantization yields the energy levels 
\be E^2_{\rm scalar} = qB\left(2n+1\right)+m^2\ \ \ \ \ n=0,1,2,\ldots\nn\ee
Now consider a fermion of charge $q$ and mass $m$. The Landau levels take the same form, but with an extra contribution analogous to Zeeman splitting of different spin states.  With gyromagnetic factor $g_e=2$, the energy levels are
\be E^2_{\rm fermion} = q\left[B\left(2n+1\right) + 2Bs\right] +m^2\nn\ee
where the spin takes values $s=+1/2, -1/2$.  For massless fermions, with $m=0$, the Zeeman splitting for spin down fermions exactly cancels the zero point energy and the resulting $n=0$   Landau level famously has vanishing energy. We will discuss the fate of such modes in the next Section.

\para
Of more immediate concern is the effect of the magnetic field on the W-bosons. These have spin 1 and charge $q=1$. One can show that the 
 gyromagnetic factor is again $g_W = 2$ and the Zeeman splitting now overcompensates for the Landau level zero point energy. The states have energy
\be
E^2_{\rm W-boson} = B\left(2n+1\right) + 2Bs + m_W^2\nn\ee
where the spin takes values $s=-1,0,+1$. When the magnetic field exceeds the critical value $B>m_W^2$, the lowest Landau level modes with $s=-1$ are tachyonic: the constant background magnetic field \eqn{nonabmag} is unstable to condensation of W-bosons. 

\para
The instability of the non-Abelian magnetic fields was first noticed in \cite{no}. Upon condensation, the W-bosons form a lattice structure, breaking continuous translational symmetry on the plane. Various properties of this lattice were explored in a number of papers. See, for example, \cite{ao0}-\cite{ao4} and \cite{proof}. 

\para
More recently, the gauge boson lattice has appeared in several situations. The same mechanism of instability is at play in the proposed condensation of rho-mesons in a strong magnetic field \cite{max1,max2,max3} and has been suggested as a phenomenological model of the QCD vacuum \cite{max4}. The gauge-boson lattice has been employed in the holographic context where it now captures the instability of a strongly correlated system in the presence of a magnetic field \cite{j1,j2}.  Moreover, it is closely related to the monopole wall configuration discussed \cite{ward} and, in the holographic context,  in \cite{bt,paul}.

\subsection{The W-Boson Lattice}\label{wbosonsec}

We will now provide a more pedestrian derivation of the tachyonic mode that we described above.  To this end, we focus on the equations of motion for the gauge fields. The Yang-Mills action can be decomposed as 
\be
S_{YM} = - \frac{1}{4e^2} \int d^3 x \ \  |D_\mu W_\nu - D_\nu W_\mu|^2 + 2m_W^2|W_\mu|^2 +  |f_{\mu\nu} - \frac i 2 (W_\mu W_{\nu}^\star - W_\nu W_\mu^\star)|^2\nn\ee
where $f_{\mu\nu} = \partial_\mu a_\nu - \partial_\nu a_\mu$ is the U(1) field strength and $D_\mu = \partial_\mu - ia_\mu$, reflecting the fact that the W-bosons have charge $+1$ under this unbroken $U(1)$.
%

\para
The equations of motion arising from this action admit the  solution $f_{12}=B$ and $W_\mu=0$. However, as described above, this solution is unstable. To see this it is sufficient to examine the linearised equations of motion. Working in $A_t=0$ gauge, these read
\be
\partial_t^2 W_x &=& -D_y (D_y W_x - D_x W_y) + iB W_y +m_W^2W_x   \nn \\
\partial_t^2 W_y  &=& - D_x (D_x W_y - D_y W_x) - iB W_x  +m_W^2W_y
\label{linw}\ee
supplemented by the Gauss' law constraint 
\be
\partial_t (D_x W_x + D_y W_y) = 0 \nn\ee
%
%
Taking suitable linear combinations $W_x\pm iW_y$, it is simple to see that the solutions to \eqn{linw} include  modes with $E^2=m_W^2-B$.  When $B>m_W^2$, these modes become tachyonic. 

\para
To describe these  modes more explicitly, we must pick a gauge. We choose Landau gauge, 
\be a_x=0\ \ \ {\rm and} \ \ \ a_y = Bx\label{landau}\ee
The tachyonic, lowest Landau level modes are then given by
\be
W_x = -iW_y =  \exp \left( - iky - \frac{B}{2} \left(x + \frac{k}{B}\right) ^2\right)\label{lllw}
\ee
These modes are labelled by the continuous parameter $k\in {\bf R}$.

\para
When these modes condense, they  break both rotational and translational invariance in the plane. The result is a lattice structure, similar to the familiar Abrikosov vortex lattice that arises in superconductors. The lattice spacing is set by the magnetic field itself.
The W-bosons form currents, screening the magnetic field near the vortex  cores while enhancing the magnetic field in the space between between the cores \cite{ao4}. (This is the opposite to a conventional superconductor).  

\para
At the onset of the instability, we can trust the linearised approximation \eqn{linw} and construct the  W-boson lattices from the lowest Landau level modes above. A one-parameter family of rhombic lattices is 
given by the ansatz $W = W_x = -iW_y$, where $W$ is the linear superposition
\be
W(x,y) = \bar W \sum_{n = -\infty}^\infty \exp \left( - \frac {i\pi n^2} 2 - \frac {2\pi i n\sqrt{B} y} \lambda - \frac B 2 \left( x + \frac{2\pi n}{\lambda \sqrt{B}}\right)^2 \right) \label{abrikosov}
\ee
Here, $\bar W$ is a constant which is undetermined at the linearised level. The condensation of W-bosons acts as a Higgs mechanism for the (until now) unbroken $U(1)\subset SU(2)$ and the associated photon gets a mass at the scale $\bar{W}$.

\begin{figure}[!h]
  \begin{center}
    \includegraphics[trim = 0.5in 0.5in 0.2in 0.7in, width=5.5in]{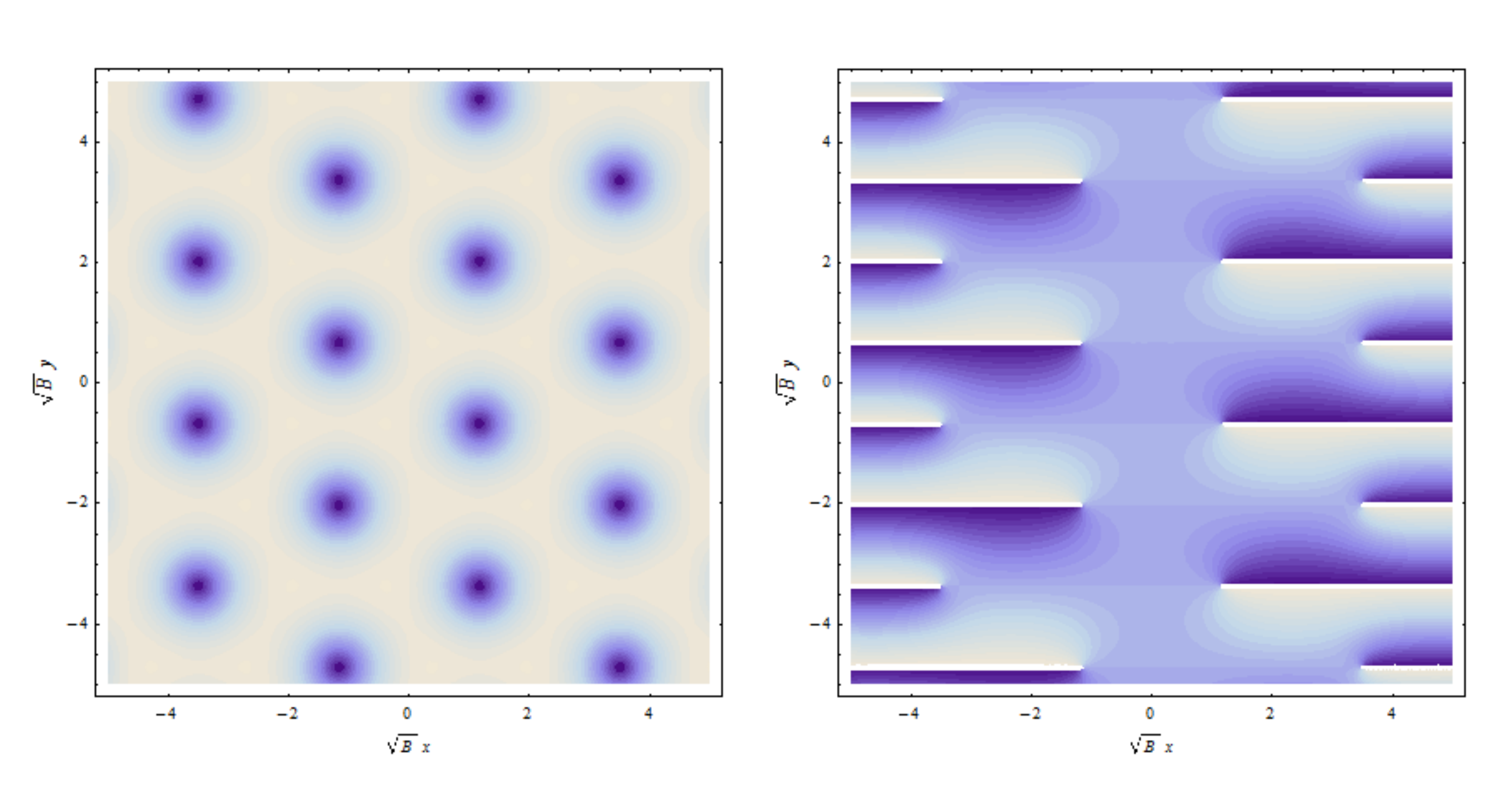}
  \end{center}
\caption{The triangular W-boson lattice. The left-hand figure shows $|W(\vec{x})|$: white means big; blue means small. The right-hand figure shows the argument of $W(\vec{x})$, defined to take values betwee $-\pi$ (blue) and $+\pi$ (white). One can see the winding of $W$ around the vortex cores.}
  \label{vortfig}
\end{figure}

\para
The parameter $\lambda \in {\bf R}^+$ determines the geometry of the lattice.  For example, a square lattice arises from the choice $\lambda^2 = {4\pi}$; a triangular lattice from  $\lambda^2 = {4\pi}/\sqrt{3}$. This can be seen by noting that the W-boson lattice \eqn{abrikosov} is invariant under the translations by the lattice vectors $\vec{d}_1 = (0,\lambda/\sqrt{B})$ and $\vec{d}_2 = (2\pi/\lambda\sqrt{B},\lambda/2\sqrt{B})$. As usual in a magnetic field, translational invariance is only assured up to phase, 
\be
W(\vec{x}+\vec{d}_1) = W(\vec{x}) \ \ \ \ {\rm and}\ \ \ \  W(\vec{x}+\vec{d}_2 ) = i\exp \left(\frac {2\pi i \sqrt{B} y } \lambda \right)\, W(\vec{x})
\nn\ee
%
%
For all values of $\lambda$, the unit cell has area $2\pi / B$ in accord with flux quantization.

\paragraph{}
The parameter $\lambda$ is arbitrary at the linearised level. It will be determined by non-linear terms which are expected to energetically prefer the triangular lattice. (The square lattice is, apparently a local, but not  global, minimum). The lowest Landau level ansatz \eqn{abrikosov} for the triangular lattice is shown in Figure \ref{vortfig}. The phase of $W(\vec{x})$ winds once around each of the vortex cores, located at $\vec{x}=(m+\ft12)\vec{d}_1+(p+\ft12)\vec{d}_2$ with $m,p\in{\bf Z}$. 
Naturally,  $W(\vec{x}) = 0$ at the core itself.

\para
The linearised approximation \eqn{abrikosov} is only justified near the onset of instability: it is no longer trustworthy when $B\gg m_W^2$.  Nonetheless, in what follows we will work exclusively with the linearised form of the lattice \eqn{abrikosov}. The reason for this is that we are ultimately interested in computing topological properties of our system. In the absence of a phase transition, these cannot change as the higher order corrections to the W-boson lattice are taken into account. In Sections \ref{fermionsec} and \ref{topsec}, we will identify a number of such properties which, although computed using the linearised lattice, remain true in the full non-linear solution. 

\para
Finally, we note that although the W-boson lattice is much studied, there remain open issues about the question of stability. Indeed, it seems unlikely that the lattice  is globally stable. We discuss some of these issues in Appendix \ref{trustapp}.

\section{Massless Fermions}\label{fermionsec}

We now turn to the fermionic excitations. We add to the action \eqn{action} two Dirac fermions, 
\be
S_{\rm Dirac} = - \int d^3x\  \ i\bar{\psi}_1\Dslash\psi_1+i\bar{\psi}_2\Dslash\psi_2\label{actionman}\ee
Both Dirac fermions $\psi_1$ and $\psi_2$ transform in the fundamental representation of the $SU(2)$ gauge symmetry. The parity anomaly prohibits an odd number of fundamental fermions \cite{redlich,agw}, which means that the pair of fermions above is the minimum number possible\footnote{The action \eqn{actionman} can be obtained by dimensionally reducing $3+1$ dimensional Yang-Mills theory coupled to a single fundamental Dirac fermion $\Psi$,
\be S = - \int d^4x\  \ i\bar{\Psi}\Dslash\Psi \nn\ee
Upon dimensional reduction, the Dirac spinor $\Psi$ becomes $\psi_1$ and $\psi_2$ above. The presence of a Dirac, as opposed to a Weyl, spinor  ensures that the theory is free from the  Witten anomaly.}.
The theory admits a $U(1)_F\times SU(2)_F$ flavour symmetry, with $\psi_i$ transforming as a doublet.

\para
As we explained above, in a constant magnetic field the $n=0$ Landau level of the fermions has zero energy. We start by describing some properties of this Landau level before moving on to study the effect of the W-boson lattice on these states\footnote{For earlier studies of fermions in fixed, background $SU(2)$ gauge fields, see \cite{wu,tan,clan}.}. 

\para
The $n=0$ Landau level has one special property: it contains only half the states of higher Landau levels. Our theory has  8 fermionic excitations. This counting arises because the Dirac spinor has two components while  our fermions are doublets of both the $SU(2)$ gauge symmetry  and the  $SU(2)_F$ flavour symmetry. While higher Landau levels are populated by all 8 species,  the $n=0$ Landau level contains only 4 species of fermions. This can be seen by explicitly solving the zero modes of the Dirac equation.  We write the $SU(2)$ gauge doublet as 
\be
\psi_i= \left(\begin{array}{c}\psi^-_{i} \\ {\psi}^+_{i}\end{array}\right)\nn\ee
where $i=1,2$ is the species index. Each component $\psi_i^\pm$ is a Dirac spinor. In Landau gauge \eqn{landau},  an orthonormal basis for the the zero modes is provided by 
\be
\psi^-_i &=&  \sqrt[4]{\frac B {2\pi} }\int  \frac{dk}{2\pi}\   \exp \left( - iky- \frac{B}{4} \left( x + \frac {2k} B \right)^2 \right) \left(\begin{array}{c}\xi^-_{i}(k) \\ 0 \end{array}\right)\label{lll1}\\
 \psi^+_i &=& \sqrt[4]{\frac B {2\pi} }\int \frac{dk}{2\pi}\   \exp \left( + iky - \frac B 4 \left( x + \frac{2k}{B} \right)^2 \right) \left(\begin{array}{c}0 \\ \xi^+_{i}(k)\end{array}\right)\label{lll2}\ee
Here $\xi_i^\pm(k)$ are four Grassmann valued collective coordinates for the $n=0$ Landau level. The argument $k\in {\bf R}$ reflects the degeneracy of this Landau level. Because the spinors transform in the fundamental representation of the $SU(2)$ gauge symmetry, the spinors  have $U(1)\subset SU(2)$ charge $\pm \frac 1 2$. This means that the degeneracy of each species is $BA/4\pi$.

\subsection{Fermions on the Lattice}

The expressions above describe the zero energy states in a constant background magnetic field. However, as we have seen, the magnetic field also causes the W-bosons to condense into a lattice structure.  We would like to understand the effect of this W-boson lattice on the fermions. Because both have the same origin, the magnetic and lattice lengths are commensurate. In particular, as fermions have charge $\pm \frac 1 2$ under $U(1) \subset SU(2)$, the area of the fermion magnetic unit cell is twice that of the W-boson lattice unit cell. 

 
\para
Substituting our expressions for the lattice \eqn{abrikosov} and fermion zero modes \eqn{lll1} and \eqn{lll2} into the action \eqn{actionman} gives us an expression for the shift in energy of the fermions, 
\be
 H  =   \sum_{i=1}^2 \int \frac{dk}{2\pi} \frac{dk'}{2\pi}  \left[ i \phi(k,k')\ \xi^{-}_i(k')^\star\,\xi_i^+(k) + {\rm h.c.}\right]\label{deltah}\ee
Here we have neglected terms coupling the $n=0$ Landau level to higher Landau levels. The effect of such terms on the energy spectrum is suppressed by $\bar{W}/\sqrt{B}$. The function $\phi(k,k')$ is a convolution of the W-boson lattice and fermionic zero modes
\be \phi(k,k') &=&\sqrt{\frac{B}{2\pi}} \int dxdy\   \exp \left( i(k'+k)y - \frac B 4 \left( x + \frac {2k} B \right)^2 - \frac B 4 \left( x + \frac {2k'} B \right)^2 \right)W(x,y)\nn\ee
%

The lattice perturbation has the effect of coupling the different $k$ modes in the $n=0$ Landau level. Our task is to diagonalise this Hamiltonian while simultaneously keeping the kinetic terms diagonal. This is achieved by the linear combination which is compatible with the magnetic translational symmetry,
\be
\zeta^+_i(p_1,p_2) &=& \frac 1 {\sqrt[4]{\pi^2 \lambda^2 B}} \sum_{n = -\infty}^\infty \exp\left(\frac{4\pi i n p_2}{\lambda\sqrt{B}}\right) \xi^+_i
\left(+p_1 + \frac{2\pi n\sqrt{B} }{ \lambda}\right)\label{zeta1}\\
\zeta^-_i(p_1,p_2) &=& \frac 1 {\sqrt[4]{\pi^2 \lambda^2 B}} \sum_{n = -\infty}^\infty \exp\left(\frac{4\pi i n p_2}{\lambda\sqrt{B}}\right)\, \xi^-_i\left(-p_1+  \frac{2\pi  n \sqrt{B}}{\lambda}\right)
\label{zeta2} \ee
%
%
%

 \begin{figure}[!h]
  \begin{center}
    \includegraphics[trim = 0.5in 0.5in 0.2in 0.7in, width=4in]{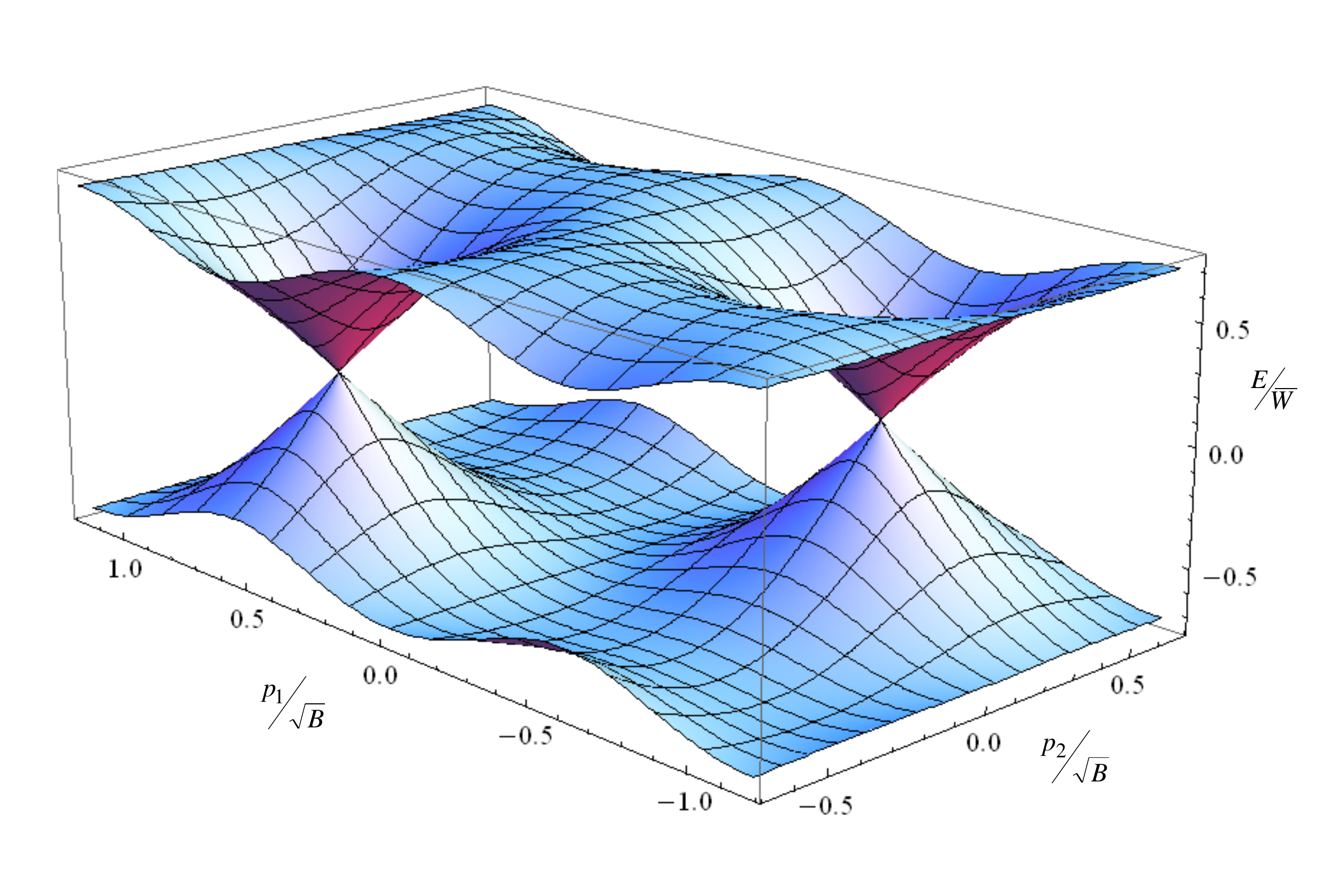}
  \end{center}
  \caption{A single Brillouin zone with its pair of Dirac cones.}
  \label{upfig}
\end{figure}

\noindent
We have replaced the $k\in {\bf R}$ index labelling the $n=0$ Landau level modes with a pair of indices $\vec{p}=(p_1,p_2)$. From the definitions, it is simple to check that these variables are periodic, taking values in the range
\be p_i\in [-\Delta p_i/2,+\Delta p_i/2)\ \ \ \ &{\rm with}&\ \ \ \ \Delta p_1 = \frac{2\pi\sqrt{B}}{\lambda} \ \ \ ,\ \ \ \Delta p_2=\frac{\lambda\sqrt{B}}{2}\label{mbz}\ee
%
%
In Appendix \ref{blochsec}, we demonstrate that $\vec{p}$ are (up to a trivial relabelling) the magnetic Bloch momenta and the torus ${\bf T}^2$ that  they  parameterize is the magnetic Brillouin zone. For now, let us simply note that the area of this torus is ${\rm Area}({\bf T}^2) = \pi B$, which follows from the fact that  the area of the magnetic unit cell  in real space is $A_\square = 4\pi/B$ and the area of the corresponding unit cell in reciprocal space should be $(2\pi)^2/A_\square =\pi B$.

 \para
In terms of the new variables $\zeta_i^\pm(\vec{p})$, the Hamiltonian \eqn{deltah} becomes
\be H = \frac{1}{\sqrt{2}} \sum_{i=1}^2\int_{{\bf T}^2} d^2p \ \left[ iW\left(\frac{2p_1}{B},\frac{2p_2}{B}\right)\, \zeta_i^{-}(\vec{p})^\star\,\zeta^+_i(\vec{p}) + {\rm h.c.} \right]\label{diagonal}\ee
A proof of this result is provided in Appendix \ref{diagonalapp}. Something rather nice has happened here. The function $W(\vec{x})$, which describes the profile of the W-boson lattice in real space, has a dual role as the energy  function $W(2\vec{p}/B)$ over the magnetic Brillouin zone.

\para
The Hamiltonian \eqn{diagonal} is diagonal in momentum modes, but still requires a trivial diagonalisation in the $\zeta^{\pm}$ index. The result is that each species $i=1,2$ gives rise to a pair of states with energy $\pm |W(2\vec{p}/B)|/\sqrt{2}$. This spectrum is drawn in Figure \ref{upfig}. In the ground state, the lower band with negative energies is filled. The excitation spectrum then consists of particles and holes, both with energy $+|W|/\sqrt{2}$. We describe the quantisation of the fermions in more detail in the next section. For now 
 note that, for each species, there exist two gapless points $\vec{p}=\vec{p}_\star$ within the magnetic Brillouin zone where $|W(2\vec{p}_\star/B)|=0$. Using the form of the linearised W-boson lattice \eqn{abrikosov}, these Dirac points are seen to lie at\footnote{A repeated word of caution: a simple relabelling is needed before interpreting these as Bloch momenta. The true Bloch momenta are $(p_x,p_y)=(p_2,p_1)$. This is explained in Appendix \ref{blochsec}.}\footnote{This result generalises to $d=3+1$. Applying a background  non-Abelian magnetic field in the $z$ direction, the W-bosons now condense to form a lattice of vortex flux tubes which are translationally invariant in $z$. Fermion propagation in the $x$-$y$ plane is governed by \eqn{diagonal}, while propagation in the $z$-direction is trivial. The result is that there are again four Dirac points with momenta
\be \vec{p}_\star = \pm \left(\frac{\Delta p_1}{4},-\frac{\Delta p_2}{4} , 0 \right) \nn \ee}
 \be \vec{p}_\star = \pm \left(\frac{\Delta p_1}{4},-\frac{\Delta p_2}{4}\right)
 \label{dpoint}\ee

\subsection{Robust Dirac Points}\label{robustsec}

The structure of fermions that we have uncovered in this section is that of a semi-metal. There are four points -- two for each flavour -- at which the spectrum is gapless, with the fermions described by a Dirac cone. The positions of these Dirac points lie at momenta \eqn{dpoint}. 
As with all our analysis, this result was derived using the linearised W-boson lattice. The purpose of this section is to show that these results are robust. They continue to hold in the full non-linear lattice. Moreover, they are unaffected by quantum corrections (at least in the absence of a phase transition).

\para
The robust nature of the Dirac points follows  the arguments given in \cite{volovikbook,wenzee,beri} (many of which of which can be traced back to \cite{nino1,volovik}). It hinges on the winding of $W(2\vec{p}/B)$, which, in real space, is shown in  Figure \ref{vortfig}. Around each of the Dirac points, the phase of $W(2\vec{p}/B)$ winds once.  These Dirac points are vortices in momentum space and, due to their topological nature, are protected against perturbations.


\para
 To see this winding in more detail, consider the transformation of 
$W(2\vec{p}/B)$ under reciprocal lattice translations. 
It is simple to check that $W$ is invariant under $p_2\rightarrow p_2+\Delta p_2$ while, under a translation along $p_1$, it picks up a phase
%
%
%
%
%
\be W\left(\frac{2(p_1+\Delta p_1)}{B},\frac{2p_2}{B}\right) = e^{4\pi i p_2/\Delta p_2}W\left(\frac{2p_1}{B},\frac{2p_2}{B}\right)\label{wwind}\ee
Computing the winding around the magnetic Brillouin zone   we have
\be
\Delta \arg  W = -i \oint  W^{-1} \nabla W = 4\pi\nn
\ee
This result, which follows solely from the translational properties of $W$, tells us that $W$ has to have zeroes: these are the two Dirac points. 

\para
One may wonder how $W$ can have a non-zero winding over ${\bf T}^2$, a  compact manifold without boundary. The answer lies in the fact that it is the Hamiltonian \eqn{diagonal} which must be single valued over the torus. The winding of $W(\vec{p})$ is cancelled by an opposite winding of the $\zeta^\pm$ operators. Indeed, the vortices in momentum space are effectively inherited from the well-known winding of magnetic Bloch states. 
The latter can be seen in our operators which, under translation, transform as
%
%
\be \zeta^+_i(p_1+\Delta p_1,p_2)&=&e^{-2\pi i p_2/\Delta p_2}\zeta^+_i(p_1,p_2)\nn\\ \zeta^-_i(p_1+\Delta p_1,p_2)&=&e^{+2\pi i p_2/\Delta p_2}\zeta^-_i(p_1,p_2)\label{zwind}\ee
In Section \ref{tknnsec}, we will see that this winding underlies the nontrivial  TKNN invariants (or first Chern numbers) of the fermion bands. 

\para
Let us now discuss how these windings persist as the W-boson lattice relaxes to its fully non-linear configuration. It can be shown that the invariance of the system under time reversal, charge conjugation, magnetic lattice translations and $U(1)_F\times SU(2)_F$ global transformations ensures that the quadratic part of the Hamiltonian takes the form
\be H =  \frac{1}{\sqrt{2}} \sum_{i=1}^2\int_{{\bf T}^2} d^2p \ \left[ iW_{NL}\left(\frac{2\vec{p}}{B}\right)\, \zeta_{i,NL}^{-}(\vec{p})^\star\,\zeta^+_{i,NL}(\vec{p}) + {\rm h.c.} \right]\label{dia}\ee
where the fermions $\zeta^{\pm}_{i,NL}$ are adiabatically connected to functions $\zeta^{\pm}_{i}(\vec{p})$ defined in \eqn{zeta1} and \eqn{zeta2} and the function  $W_{NL}(\vec{p})$ is smoothly connected to the function $W(\vec{p})$ defined in \eqn{abrikosov}. Furthermore, this form must also hold in the presence of loop corrections.

\para
In the absence of a phase transition, the winding of $W_{NL}$ cannot change as the W-boson lattice relaxes to its non-linear configuration. The pair of Dirac points for each species of fermion persist.

\subsubsection*{Location of the Dirac Points}

We will now show that, given the existence of two Dirac points for each fermion, the symmetries are also sufficient to dictate where they sit. 
We will need to make a couple of technical assumptions which boil down to the requirement that $W_{\rm NL}$ inherits certain discrete symmetries from  from $W(\vec{p})$. The first is translational symmetry. The linearised lattice obeys
\be W\left(\frac{2p_1}{B} + \frac{\Delta p_1}{B}, \frac{2p_2}{B}+\frac{\Delta p_2}{B}\right) = i e^{2\pi i p_2/\Delta p_2}\,W\left(\frac{2p_1}{B},\frac{2p_2}{B}\right)\label{wtrans}\ee
The existence of this symmetry can be traced to the fact that the lattice unit cell (in real space) has only half the area of the magnetic unit cell. We assume that the function $W_{NL}$ continues to transform covariantly under translations. The second is an inversion symmetry
\be W\left(\frac{2\vec{p}}{B}\right) = W\left(-\frac{2\vec{p}}{B}\right)\label{winv}\ee
We assume that $W_{NL}$ also enjoys this property.  

%
%
%
%

\para
The translational symmetry \eqn{wtrans} and inversion symmetry \eqn{winv}  ensure that the Dirac points, defined by $W_{NL}(2\vec{p}_\star/B)=0$, sit at one of two positions: either
\be
\vec{p}_\star = \pm \left(\frac{\Delta p_1}{4},-\frac{\Delta p_2}{4}\right)\nn\ee
or 
\be \vec{p}_\star =\pm  \left(\frac{\Delta p_1}{4},\frac{\Delta p_2}{4}\right)\nn\ee
Comparing to \eqn{dpoint}, we see that our lattice $W$ realises the first of these possibilities. This choice cannot  change as the lattice is smoothly deformed\footnote{While this statement is true, in Appendix \ref{blochsec} we will see that there is an ambiguity in the definition of the choice of Bloch momenta (essentially a choice of origin) which means that the physics is left unchanged for lattices which realise the second choice above.}.

\para
To conclude this section, let us finish by describing a property which is not robust to corrections: the speed of propagation of the Dirac modes. For the linearised lattice \eqn{abrikosov}, it is clear that the dispersion relation near the Dirac point takes the form $E=c|\vec{p}|$ where the speed is given by $c\sim |\bar{W}|/\sqrt{B}\ll 1$. As the ratio $B/m_W^2$ increases, the speed of propagation is expected to increase.

\section{Massive Fermions}\label{topsec}

In this section, we would like to explore in more detail the structure of the fermionic modes. To do so, we first open up a gap in the spectrum. This is achieved by simply returning to our original action \eqn{actionman} and adding a mass for the fermions. We choose to give the different species $\psi_1$ and $\psi_2$ equal and opposite masses, 
\be S_{\rm mass} = \int d^3x\ im(\bar{\psi}_1\psi_1-\bar{\psi}_2\psi_2)\label{smass}\ee
The significance of this choice lies in the observation that it preserves time reversal, a fact that will prove to be important later. To see this, note that in $d=2+1$ dimensions the fermionic bilinear $\bar{\psi}\psi$ is odd under time reversal. However, we can define a new version of this symmetry in which we simultaneously exchange $\psi_1$ and $\psi_2$. The operator \eqn{smass} is invariant under this new symmetry and it also preserves charge conjugation. Full details of the action of time reversal invariance as well as charge conjugation are provided  in Appendix \ref{dissec}.

\para
The addition of the mass term breaks the $SU(2)_F$ flavour symmetry down to a $U(1)_A$ subgroup. Under the pair of Abelian symmetries $U(1)_F\times U(1)_A$, $\psi_1$ has charge $(+1,+1)$ while $\psi_2$ has charge $(+1,-1)$.

\subsection{Quantisation}\label{quantsec}

We now look a little closer at the quantum states in the $n=0$ Landau level. The effect of the mass  is, unsurprisingly, to open up a gap in the spectrum. In terms of the Bloch momenta, the Hamiltonian becomes
\be  H =  \int_{\rm {\bf T}^2} d^2p &\Big[& m\zeta_1^{-\star}\zeta_1^- - m\zeta_1^{+\star}\zeta_1^+ +i \frac{W}{\sqrt{2}}\zeta_1^{-\star}\,\zeta^+_1 -i\frac{W^\star}{\sqrt{2}}\zeta_1^{+\star}\zeta_1^- \nn\\
&\ \ & -m\zeta_2^{-\star}\zeta_2^- + m\zeta_2^{+\star}\zeta_2^+ +i \frac{W}{\sqrt{2}}\zeta_2^{-\star}\,\zeta^+_2 -i\frac{W^\star}{\sqrt{2}}\zeta_2^{+\star}\zeta_2^-\Big]\nn\ee
 \begin{figure}[!h]
  \begin{center}
    \includegraphics[trim = 0.5in 0.5in 0.2in 0.7in, width=4in]{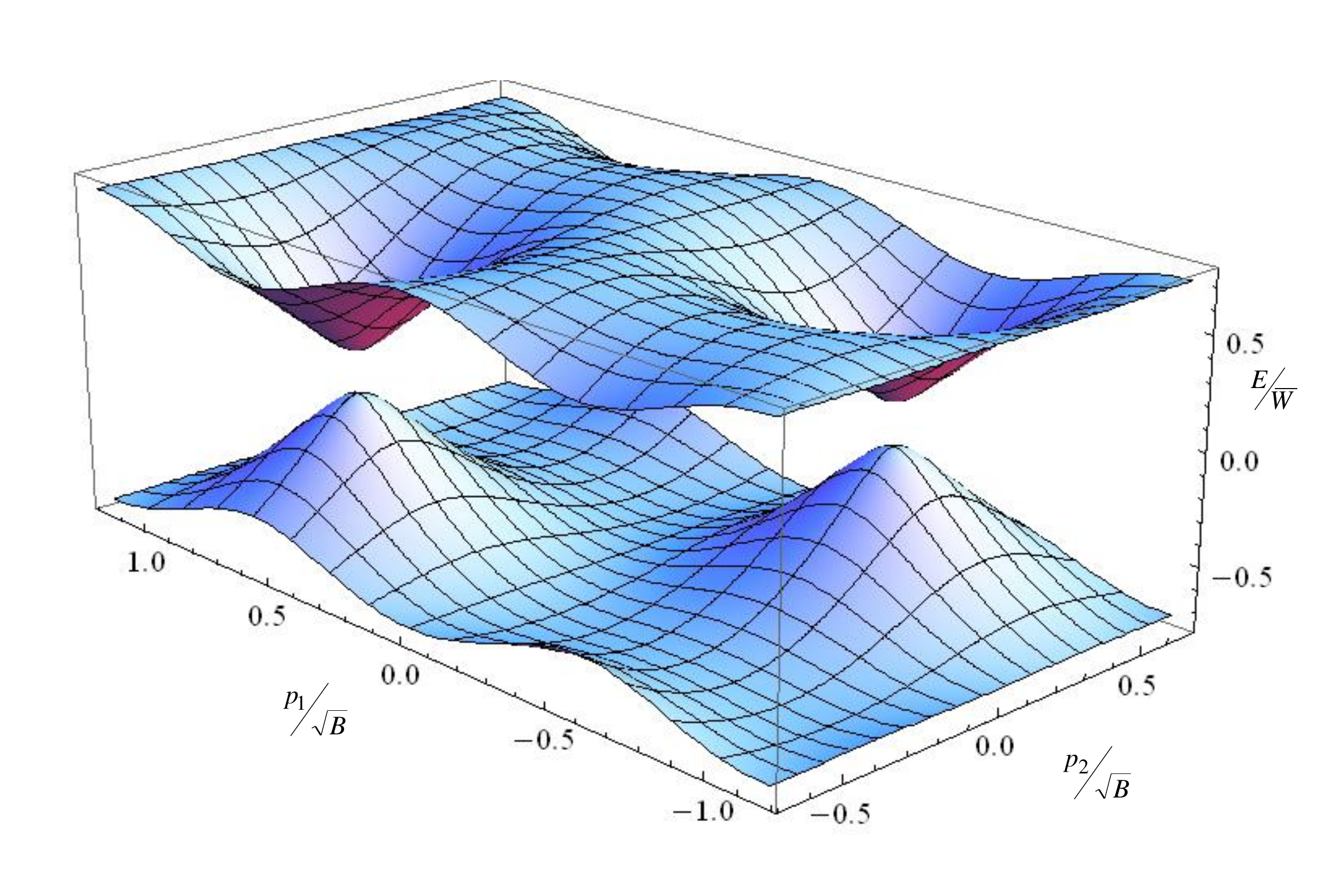}
  \end{center}
  \caption{A single, gapped Brillouin zone.}
\label{bitofgap}
\end{figure}

\noindent
Quantising the fermions gives rise to a pair of particles and a pair of anti-particles, each with energy
\be E(\vec{p})^2 = m^2 + \frac{1}{2}|W(2\vec{p}/B)|^2\nn\ee
The resulting band structure is shown in the figure. 
To look at these states in more detail, we diagonalise the $\zeta_i^\pm$ components. This is simply achieved by introducing  annihilation and creation operators $a_i,a_i^\star$ and $b_i,b_i^\star$ with $i=1,2$. Assuming that $m>0$, we write the creation operators as
\be
a_1^\star & = & \sqrt{\frac{E+m}{2E}} \zeta_1^{-\star} - \frac{iW^\star}{\sqrt{4E(E+m)}} \zeta_1^{+\star}  \nn \\ 
b_1^\star & = & \frac{iW^\star}{\sqrt{4E(E+m)}} \zeta_1^-  +  \sqrt{\frac{E+m}{2E}} \zeta_1^+  \nn \\
a_2^\star & = & \frac{iW}{\sqrt{4E(E+m)}} \zeta_2^{-\star} +  \sqrt{\frac{E+m}{2E}} \zeta_2^{+\star}  \nn \\
b_2^\star & = & \sqrt{\frac{E+m}{2E}} \zeta_2^{-} - \frac{iW}{\sqrt{4E(E+m)}} \zeta_2^{+}  \label{creationops}
\ee
%
%
%
where $E=\sqrt[+]{E^2}$. (Strictly speaking, the collective coordinates $\zeta$ should be evaluated at $t=0$ in the above expressions).  This choice of basis has the advantage that it remains non-degenerate at the points $W=0$. It is simple to check that the canonical equal time anti-commutation relations for $\psi$ or, equivalently, $\zeta$, translate into the canonical anti-commutation relations for $a$ and $b$,
\be \{a_i(\vec{p}),a^\star_j(\vec{p}\,') \} = \delta_{ij}\,\delta(\vec{p}-\vec{p}\,')\ \ \ \ {\rm and}\ \ \ \ \ \{b_i(\vec{p}),b^\star_j(\vec{p}\,')\} = \delta_{ij}\,\delta(\vec{p}-\vec{p}\,')\nn\ee
In terms of these new operators, the lowest Landau level Hamiltonian is, after normal ordering, simply
\be
 H = \sum_{i=1}^2 \int_{\rm BZ}d^2p\ \ E(a_i^\star a_i +  b_i^\star b_i) 
\ee
The ground state $|\Omega\rangle$ of the system obeys $a_i|\Omega\rangle = b_i|\Omega\rangle =0$ and 
is neutral under the $U(1)_F\times U(1)_A$ global symmetry. 

\para
The excited states fall into one of four bands: a pair of particles and a pair of anti-particles. Each is labelled by the Bloch momenta $\vec{p}\in {\rm BZ}$ and carries flavour charge,
\begin{center}
\begin{tabular}{c|cccc}
& $a^\star_1|\Omega\rangle$ & $b^\star_1|\Omega\rangle$ & $a^\star_2|\Omega\rangle$ & $b^\star_2|\Omega\rangle$  \\
\hline
$U(1)_F$ & +1 & $-1$ & +1 & $-1$ \\
$U(1)_A$ & +1 & $-1$ & $-1$ & $+1$ \\
\end{tabular}\end{center}
However, there are further quantum numbers that can be associated to each band that follow from topological considerations. In the next two sections we turn to these, first dealing with the TKNN invariants and then moving on to the more subtle $\mathbb{Z}_2$ invariant.

\subsection{TKNN  Invariants}\label{tknnsec}

In the presence of a gap, bands can be classified by a number of topological invariants. These arise from the winding of the states as we move around the Brillouin zone ${\bf T}^2$. The simplest of these invariants is the first Chern number or TKNN invariant \cite{tknn}. 

\para
We start by focussing on the band of states $a_1^\star(\vec{p})|\Omega\rangle$. As we move around the magnetic Brillouin zone, parameterised by $\vec{p}$, the phase of these states changes.  This is captured by the Berry connection, an Abelian 1-form over ${\bf T}^2$ defined by
\be 
\gamma(\vec{p})  = i\langle \Omega| a_1\frac{\partial}{\partial \vec{p}}\,a_1^\star|\Omega\rangle\,\cdot d\vec{p}\nn\ee
The TKNN invariant is then defined to be the first Chern number,
\be C_1 = \frac{1}{2\pi}\int_{{\bf T}^2}\,d\gamma\nn\ee
Analogous definitions also hold for the other three bands of states, $a_2^\star \vert \Omega \rangle$, $b^\star_1 \vert \Omega \rangle$ and $b^\star_2 \vert \Omega\rangle$.

\para
To evaluate the first Chern number, we need to know how the states transform under translations through reciprocal lattice vectors $\Delta p_1$ and $\Delta p_2$ defined in \eqn{mbz}. We assume that the ground state $|\Omega\rangle$ is invariant. One can check that the  creation operators defined in \eqn{creationops} are invariant under $p_2\rightarrow p_2+\Delta p_2$. However, under a translation along $p_1$, they pick up a phase. 
\be
a_1^\star(p_1+\Delta p_1,p_2) &=& e^{-2\pi i p_2/\Delta p_2} \,a_1^\star(p_1,p_2) \nn\\ a_2^\star(p_1+\Delta p_1,p_2)&=& e^{+2\pi i p_2/\Delta p_2} \,a_2^\star(p_1,p_2)\nn\\ 
b_1^\star(p_1+\Delta p_1,p_2) &=& e^{-2\pi i p_2/\Delta p_2} \,b_1^\star(p_1,p_2) \nn\\ b_2^\star(p_1+\Delta p_1,p_2)&=& e^{+2\pi i p_2/\Delta p_2} \,b_2^\star(p_1,p_2) \label{abtrans} \ee
%
%

\begin{figure}[!h]
  \begin{center}
    \includegraphics[trim = 0.5in 0.5in 0.2in 0.1in, width=3in]{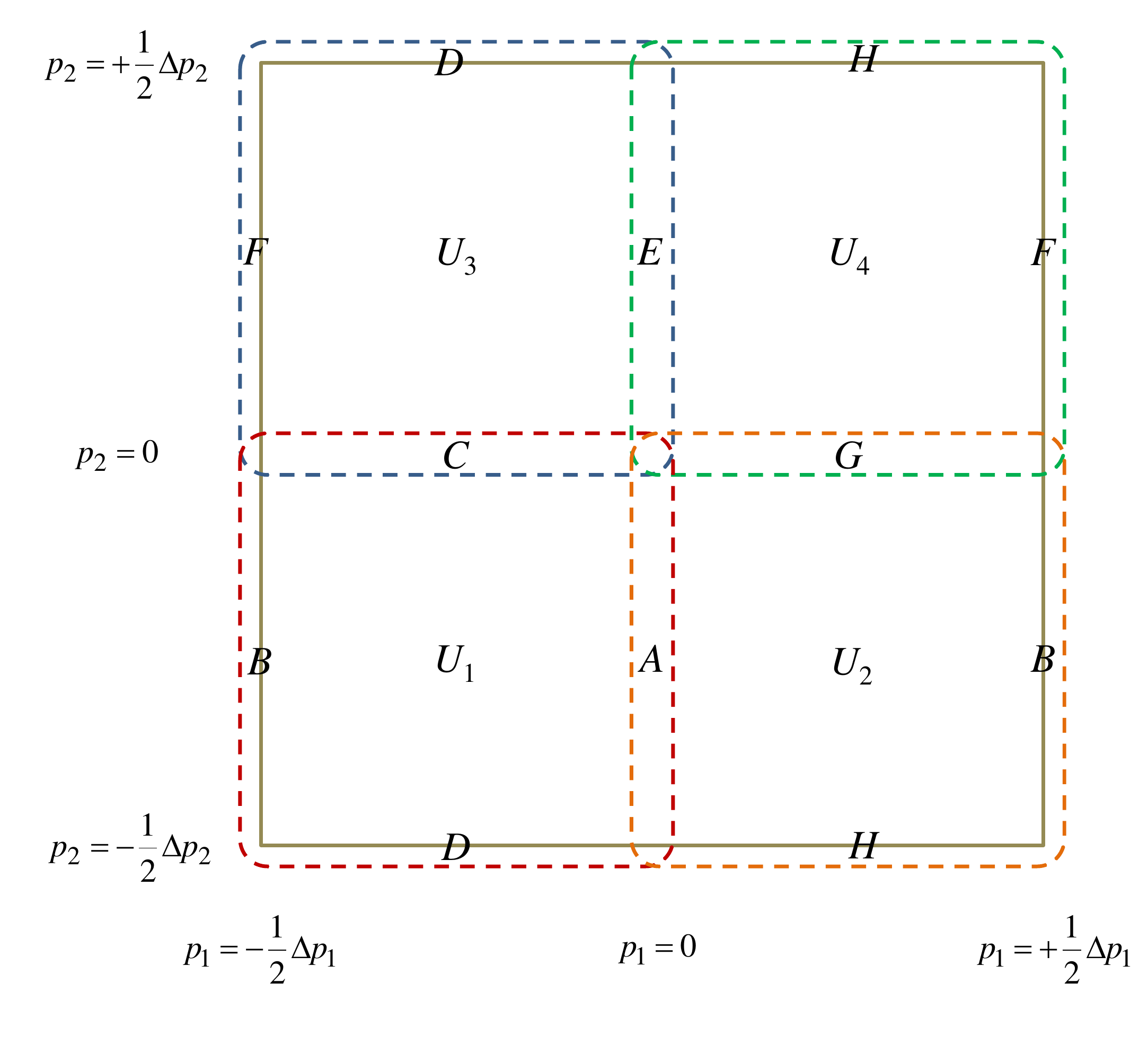}
  \end{center}
  \caption{Patches over the magnetic Brillouin zone.}
  \label{patchfig}
\end{figure}

To proceed, we cover the magnetic Brillouin zone with four charts, $U_a$, $a=1,2,3,4$. These charts are defined in Figure \ref{patchfig} where we have also given names $A$ through $H$ to a number of overlap regions.

\para
By Stokes' theorem, we can express the Chern number as a sum of integrals of the connection around the boundaries of the four charts.
\be
C_1 = \frac 1 {2\pi}  \left( \oint_{\partial U_1} \gamma + \oint_{\partial U_2} \gamma + \oint_{\partial U_3} \gamma +\oint_{\partial U_4} \gamma \right)
\nn\ee
Naively one might think that this vanishes vanishes because we're integrating along each edge twice, once in each direction. However, the fact that the creation operators transform by a phase under translations by reciprocal lattice vectors means that the Berry connections evaluated on the edges of different charts differ by a gauge transformation. This is responsible for the non-vanishing Chern number. 

\para
Let's focus once again on the $a_1^\star|\Omega\rangle$ states. In overlap region $B$, we have 
\be
\left.\gamma_{p_2}\right|_{U_2} = \left.\gamma_{p_2}\right|_{U_1}+\frac{2\pi}{\Delta p_2}\nn\ee
Similarly, in overlap region $F$, we have
\be
\left.\gamma_{p_2}\right|_{U_4} = \left.\gamma_{p_2}\right|_{U_3}+\frac{2\pi}{\Delta p_2}\nn\ee
On all other overlap regions, the transformation properties are trivial.  The resulting TKNN invariant for the $a_1^\star|\Omega\rangle$ band is therefore
\be 
C_1  = \frac 1 {2\pi} \int_B \left( \left.\gamma_{p_2}\right|_{U_2} - \left.\gamma_{p_2}\right|_{U_1}\right) dp_2 +  \frac 1 {2\pi} \int_F \left( \left.\gamma_{p_2}\right|_{U_4} - \left.\gamma_{p_2}\right|_{U_3}\right) dp_2 =+1 \nn\ee
Very similar calculations hold for the other bands. We can now extend the table listing the $U(1)_F$ and $U(1)_A$ flavour charges of the different bands to also include the Chern number,
\begin{center}
\begin{tabular}{c|cccc}
& $a^\star_1|\Omega\rangle$ & $b^\star_1|\Omega\rangle$ & $a^\star_2|\Omega\rangle$ & $b^\star_2|\Omega\rangle$  \\
\hline
$U(1)_F$ & $+1$ & $-1$ & $+1$ & $-1$ \\
$U(1)_A$ & $+1$ & $-1$ & $-1$ & $+1$ \\
$C_1$ & $+1$ & $+1$ & $-1$ & $-1$\\
\end{tabular}\end{center}
%

\para
It is worth seeing how these Chern numbers fit in with the discrete symmetries defined in Appendix \ref{dissec}. Under charge conjugation, the states $a_1^\star|\Omega\rangle$ map into $b_1^\star|\Omega\rangle$. Correspondingly, both have the same Chern number. Meanwhile, under time reversal invariance the states $a_1^\star|\Omega\rangle$ map into $a_2^\star|\Omega\rangle$. This is reflected in the fact that they carry opposite Chern number. 

\para
These TKNN invariants can be related to the Dirac points that arise when the gap closes. We saw in Section \ref{robustsec} that the winding of the phase acquired by the fermions as they traverse the Brillouin zone \eqn{zwind} is directly related the phase acquired by the W-bosons \eqn{wwind}. Moreover, the phase acquired by the W-bosons is responsible for the pair of Dirac points that lie in the Brillouin zone. The upshot is that the Chern numbers in the gapped phase are directly related to the number of Dirac points in the gapless phase. Specifically, for each species of fermion,
\be \hash\,\mbox{Dirac points} = |C_1(a_i^\star|\Omega\rangle) + C_1(b_i^\star|\Omega\rangle)| = 2\nn\ee

\subsection{$\mathbb{Z}_2$ Invariants}

In a theory with a time reversal symmetry ${\cal T}$ with the property that ${\cal T}^2 =-1$, there is a more subtle topological invariant that one can compute: this is the $\mathbb{Z}_2$ invariant $\nu$ \cite{km,fukanepump,moorebalents,fk}.

\para
We describe the time reversal invariance of our system in Appendix \ref{dissec}. There are, in fact, two different time reversal symmetries that one can define. We call these $\tilde{\cal T}$ with $\tilde{\cal T}^2=+1$, defined in in \eqn{t}, and $\tilde{\cal T}^\prime$ with $\tilde{\cal T}^{\prime\,2}=-1$, defined in \eqn{tprime}. It is the existence of the latter that allows us to define the $\mathbb{Z}_2$ invariant in our system.

\para
The $\mathbb{Z}_2$ index $\nu$ characterises pairs of bands related by time reversal. For our system, we can associate a $\mathbb{Z}_2$ index to the $a_i|\Omega\rangle$ pair and another $\mathbb{Z}_2$ index to the $b_i^\star|\Omega\rangle$ pair. While $\nu$ is usually difficult to compute, it can be easily obtained when a $U(1)_A$ symmetry is present. In this case, it is simply  the parity of the Chern number of either of the two bands within the pair \cite{shengspinchern,rahulspinZ2,fukuispinZ2}
\be \nu=C_1\ \ \ \  \textrm{mod}\  2\label{z4}\ee 
For our system, this gives
\be \nu(a_i|\Omega\rangle)=\nu(b_i^\star|\Omega\rangle)=1 \nn \ee
%
The same result can also be obtained using the inversion symmetry of the theory. This is discussed in Appendix \ref{z2sec}.

\para
The above $\mathbb{Z}_2$ invariants characterise pairs of bands. Usually, however, one is interested in the $\mathbb{Z}_2$ invariant for the whole system. The relationship is straightforward: one simply sums over pairs of filled bands
\be \nu_{\rm tot}=\sum_{j \in\ \rm filled\ pairs}\nu_j\  \ \ \ \ \ \ \textrm{mod}\  2\label{eq:Z2tot}\ee
In order to make sense of which bands are filled and which are unfilled, we need to return to the 1920's and a Dirac sea picture of the fermions, in which the anti-particles are holes in filled negative energy states. To this end, we introduce the ``zero-energy" state, 
\be |0\rangle = \prod_{\vec{p}} b_1^\star (\vec{p})b_2 ^\star (\vec{p})|\Omega\rangle\nn\ee
Then the table of the different bands of excitations can be rewritten as
\begin{center}
\begin{tabular}{c|cccc}
& $a^\star_1|0\rangle$ & $b_1|0\rangle$ & $a^\star_2|0\rangle$ & $ b_2|0\rangle$  \\
\hline
$U(1)_F$ & $+1$ & $+1$ & $+1$ & $+1$ \\
$U(1)_A$ & $+1$ & $+1$ & $-1$ & $-1$ \\
$C_1$ & $+1$ & $-1$ & $-1$ & $+1$\\
\end{tabular}\end{center}
In the $|\Omega\rangle$ vacuum, the time-reversed pair of bands $b_i|0\rangle$ is filled. Using the definition of the $\mathbb{Z}_2$ index in terms of the Chern number \eqn{z4}, we have
$\nu(b_i|0\rangle)=1$.  In the same picture, higher Landau level antiparticle states are also all filled. However, these higher Landau levels have twice the states of the $n=0$ level. This means that  they consist of two time-reversed pairs of  bands, each with $\nu=1$,  and  do not contribute to the total sum (\ref{eq:Z2tot}). We thus have
\be\nu_{\rm tot}=1 \ee
strongly suggesting that the Dirac-Yang-Mills theory is a topological insulator. We will return to this conclusion below.

\para
From the discussion above, it appears that there is no more information in the $\mathbb{Z}_2$ invariant than in the original TKNN invariants. This is not quite true. We will review why the $\mathbb{Z}_2$ invariant is the more robust than the TKNN invariants in Section \ref{sumsec}.

\subsection{Quantum Spin-Hall Effect}

The existence of  a non-trivial $\mathbb{Z}_2$ invariant in a theory with a $U(1)_F\times U(1)_A$ global symmetry is usually associated with the quantum spin-Hall effect  \cite{kanemele,bernevigzhang}. Here the $U(1)_F$ global symmetry is identified with electromagnetism while the ``spin" is the conserved charge associated to the  $U(1)_A$ symmetry. We now  discuss how the quantum spin-Hall effect arises in the Dirac-Yang-Mills theory.

\para
Both the Hall conductivity and the spin-Hall conductivity are directly related to the TKNN invariants. In our system, the Hall conductivity for both $U(1)_F$ and $U(1)_A$ vanishes, a fact which is guaranteed by time reversal symmetry. However, a mixed Hall conductivity survives. This is the spin-Hall effect: a background electric field for $U(1)_F$ results in a ``spin" current for $U(1)_A$.  The TKNN formula \cite{tknn,haldane} for the spin-Hall conductivity 
is\footnote{There is a slight subtlety here associated to the difference between particle physics and condensed matter language. In condensed matter, the sum over filled bands of the Fermi sea is always finite. In the high-energy language, the sum over bands in the Dirac sea is infinite. A slightly better formulation in the high-energy context, in  form that is manifestly invariant under charge conjugation,  is  \be
\sigma_{xy} = -\frac{e^2}{4\pi} \left( \sum_{\rm{filled \ bands}} Q_A Q_F C_1\, \,- \!\!\sum_{\rm{unfilled \ bands}} Q_A Q_F C_1 \right) \ee For our system, this agrees with \eqn{TKNN}.}
\begin{equation}
\sigma_{xy} = -\frac{e^2}{2\pi} \sum_{\rm{filled \ bands}} Q_A Q_F C_1 \label{TKNN}
\end{equation}
where $Q_A$ and $Q_F$ are the charges under $U(1)_A$ and $U(1)_F$. In our theory, this is an infinite sum  due to the filled higher Landau levels for anti-particle states.
However, as in the calculation of the $\mathbb{Z}_2$ invariant, the contribution from these states vanishes. This is because the higher Landau levels have twice the states of the $n=0$ level and, for each $Q_A$ sector, these carry opposite TKNN invariants. 
%
%
 Hence, the only non-vanishing contributions come from $b_i|0\rangle$ which give
\be
\sigma_{xy} = \frac{e^2}{\pi} \label{sigmaFA}\label{hall}\ee

While it is nice to see the derivation of the spin-Hall conductivity through TKNN invariants, we should stress that, despite appearances,  the existence of the lattice played little role in the above discussion. 
%
To highlight this, we note that the spin-Hall conductivity can be much more easily derived (at least from a particle physicists perspective) without mention of the lattice or background magnetic field. We need simply couple background gauge fields  $A_\mu^{(A)}$ and $A^{(F)}_\mu$ to the $U(1)_A$ and $U(1)_F$ currents respectively. The mass term for the fermions prevents the appearance of zero energy states as these gauge fields are introduced. Integrating out the massive fermions then results in mixed Chern-Simons coupling (see, for example, \cite{dunne} for a review),
\be L_{\rm CS} = \sigma_{xy} \,A^{(A)}\wedge F^{(F)}.\nn\ee
This is the low-energy effective action for the spin-Hall effect. A standard calculation reproduces \eqn{hall}.

\subsection{Topological Classification and Boundary Modes} 

In the previous sections, we have seen that our system carries non-zero $\mathbb{Z}_2$ invariant and exhibits the quantum spin-Hall effect. We might be tempted to conclude that we have ourselves a topological insulator of the simplest $\mathbb{Z}_2$ type (see, for example \cite{hk,zreview}). In fact, we shouldn't be too hasty. In this section we explain why it's a little premature to refer to our system as a topological insulator. In the next, we describe the deformation that is necessary to make this claim.

\para
There is a well-known classification of topological insulators based on the discrete symmetries of the problem. 
This is, at heart, a classification of Hamiltonians of free fermions  \cite{altland,heinz}, and forms the basis of the ``periodic table of topological insulators" \cite{schnyder,kitaev,ryu}.  It is natural to ask: where in this classification table does our Dirac-Yang-Mills theory sit?

\para
In Appendix \ref{dissec} we describe the discrete symmetries of our theory that leave the background W-boson lattice invariant\footnote{There is a subtle difference in the language used in high-energy physics and condensed matter physics when describing discrete symmetries. We provide a translation service in Appendix \ref{dissec}.}. The theory enjoys a charge conjugation symmetry that we call $\tilde{\cal C}$.  Importantly, the presence of the W-boson condensate means that this must be defined such that $\tilde{\cal C}^2 =-1$. There is also time reversal invariance. The existence of the $U(1)_A$ global symmetry means that there are two possibilities for time reversal: $\tilde{\cal T}$ with $\tilde{\cal T} = -1$ or $\tilde{\cal T}'$ with $\tilde{\cal T}'^2=-1$.

\para
The existence of these symmetries place the fermions in  Yang-Mills theory in either symmetry class CI or CII.  According to \cite{schnyder,kitaev,ryu} neither of these classes can realise topologically nontrivial massive fermionic phases in $d=2+1$ dimensions. Yet, as we have just seen, our theory has a non-vanishing $\mathbb{Z}_2$ invariant and exhibits quantum spin-Hall physics. How to reconcile this with its position in the periodic table?

\para
The resolution lies in the observation that the adjective ``topologically nontrivial" is meaningful only if there is a ``topologically trivial" reference system; the classification table  only counts the number of topologically distinct phases without singling one of them out as ``trivial". The table merely tells us that there is only one phase in class CI/CII and $2+1$ dimensions. From our results above, we conclude that any form of a Dirac-Yang-Mills system must have a nontrivial $\mathbb{Z}_2$ index, $\nu=1$, as long as charge conjugation and time reversal symmetries are preserved. 
\para
A similar issue arises if we focus on just one of the $U(1)_A$ sectors, say $Q_A=1$. This sector alone has charge conjugation,  but not time reversal symmetry placing it in class $C$. The classification scheme of \cite{ryu} states that the Chern number should lie in $2\mathbb{ Z}$. However, this should not be interpreted to mean that $C_1$ is even; indeed, the explicit calculation above shows that $C_1=\pm 1$. Rather, it tells us that Chern numbers of topologically distinct phases differ by an even integer.

\para
The upshot of this is that  we may have  a ``non-trivial" topological insulator, but this only really makes sense if there is a ``trivial" insulator to compare with! In the next section we will describe the deformation of our theory that is necessary to realise this. First, however, let us explain how one can see these results from a slightly different perspective.

\subsubsection*{The View from the Boundary}

Many of the results from the story of topological insulators can be most simply derived by looking at the modes that propagate on the boundary between two phases. This also provides a useful viewpoint on the discussion above. 

\para
The boundary modes are intimately related to the different topological characterisations of the bulk, a relationship which sometimes goes by the name ``bulk-boundary correspondence". This correspondence states:
\begin{itemize}
\item For a domain wall interpolating  between regions with different $\mathbb{Z}_2$ indices, there are an odd number of counter-propagating pairs of modes  which cannot be given a mass if time reversal invariance is preserved. 
\item For a domain wall interpolating between regions with Chern numbers $C_1=N_1$ and  $C_1=N_2$, the difference between the number of left- and right-moving modes is $N_1-N_2$\footnote{Here, we are using the Dirac sea picture of quantum field theory. The Chern numbers that we refer to are the Chern numbers of the filled bands.}.
\end{itemize}
For more details see, for example, \cite{hk,zreview,volovikbook}. 
\para
Let us consider a system that respects charge conjugation and time reversal symmetries. Suppose that this system has a domain wall separating two distinct regions, and suppose that this domain wall supports massless propagating excitations. The dynamics of the fermions within the domain wall are described by a $1+1$ dimensional low energy action of the form
\begin{equation}
S_{1D}=\int dt dx\ i\psi^\dagger \partial_t \psi - \int dt dx \ \psi^\dagger(x) (-i V)\partial_x \psi(x), 
\end{equation}
with a Hermitian velocity matrix $V$.

\para
The discrete symmetries impose certain restrictions on the boundary modes. To see how they arise, let us examine the transformation properties of the velocity matrix $V$ under discrete symmetries, starting with charge conjugation. Charge conjugation induces the relation
\be
\hat{\mathcal{C}}\left[(-iV)\partial_x\right]\hat{\mathcal{C}}^{-1}=(iV)\partial_x
\nn\ee
where $\hat{\mathcal{C}}$ is anti-unitary\footnote{The statement that $\hat{\cal C}$ is anti-unitary is probably familiar to condensed matter theorists and foreign to particle theorists. We explain the difference in language in Appendix \ref{dissec} where we also give the concrete form of $\hat{\cal C}$ for our problem.} and $\hat{\mathcal{C}}^2=-1$. Equivalently,  
\be
\hat{\mathcal{C}}V\hat{\mathcal{C}}^{-1}=V 
\nn\ee
Using the same argument that results in Kramers' theorem, one can deduce that the velocities, which are the eigenvalues of $V$, are twofold degenerate. The number of right-moving modes must  be even. The number of left-moving modes is also even.  Using the bulk-boundary correspondence described above, this is equivalent to the statement that we saw above: 
the Chern numbers of charge conjugation invariant systems can  only differ by an even amount. In particular, the fact that each $U(1)_A$ sector of our system has $C_1=\pm 1$ tells us that, as long as we preserve $\tilde{\cal C}^2=-1$, we cannot construct a ``trivial" Dirac-Yang-Mills $U(1)_A$ sector with $C_1=0$.

\para
Time reversal, $\hat{\cal T}$,  also imposes restrictions on the boundary modes. This is again an anti-unitary operator such that
\begin{equation}
\hat{\mathcal{T}}V\hat{\mathcal{T}}^{-1}=-V
\end{equation}
Regardless of the value of $\hat{\cal T}^2$, the existence of a time reversal symmetry tells us that, unsurprisingly, the velocities come in opposite pairs:  a right-moving mode will always have a left-moving partner. 

\para
In summary, charge conjugation ensures an even number of right-movers, while time reversal invariance ensures that these are accompanied by the same number of left-movers. It is not possible to have an odd number of counter-propagating modes on the domain wall. Therefore, any domain wall can only interpolate between phases with the same $\nu$. 
This means that although our $\mathbb{Z}_2$ invariant is $\nu=1$, usually thought to be ``non-trivial", the existence of charge conjugation prevents a  ``trivial" phase with $\nu=0$.

\subsection{Breaking Charge Conjugation: A Topological Insulator}

We have seen that, although our bands have non-trivial $\mathbb{Z}_2$ invariant, any deformation of this system that preserves the charge conjugation symmetry also has non-trivial $\mathbb{Z}_2$ invariant. That's a little underwhelming: interesting things happen only when we can place two systems with different $\mathbb{Z}_2$ invariants in contact.

\para
To realise a deformation that carries a different $\mathbb{Z}_2$ invariant, it is clear what we must do: break charge conjugation. The simplest such operator is a chemical potential $\mu$ for the $U(1)_F$ flavour symmetry\footnote{One can also break charge conjugation by introducing a Yukawa coupling $\bar{\psi}\phi\psi$ between the fermions and Higgs field.},
\be H =  \mu \sum_{i=1,2} \int d^2 x \ \bar{\psi}_i\gamma^0\psi_i\nn\ee
This clearly breaks charge conjugation since, for $|\mu| > m$, we begin to fill (or deplete) one of the bands. However, it preserves time reversal invariance. At this point, the distinction between our two time reversals, $\tilde{\cal T}$ in \eqn{t} and $\tilde{\cal T}^\prime$ in \eqn{tprime} becomes important again. It is the presence of $\tilde{\cal T}^\prime$ that is most crucial, since it ensures that our system enjoys Kramers degeneracy, due to the fact that $\tilde{\cal T}^2 =-1$.
In terms of the classification table, this places us in class AII, where we have a $\mathbb{Z}_2$ grading of insulating states in $2+1$ dimensions. Indeed, this is the class that houses the original quantum spin-Hall effect.

\para We have already seen that it is possible to realise a phase in which the $\mathbb Z_2$ invariant is non-trivial, with $\nu = 1$. It remains to find a phase in which the $\mathbb{Z}_2$ invariant is trivial, with $\nu=0$. To achieve this, we tune the chemical potential to lie in the gap between the $n=0$ Landau level and the higher Landau levels. This means that  the $a_i^\star|0\rangle$ bands are filled for both $i=1,2$. (These are the bands shown in the top of Figure \ref{bitofgap}). We can now revisit our calculation of the $\mathbb{Z}_2$ invariant given in \eqn{eq:Z2tot}. With the extra bands filled, we have
\be \nu_{\rm tot}=C_1(b_i|0\rangle) + C_1(a_i^\star|0\rangle)\  \textrm{mod}\  2 \nn=0\ee 
The same conclusion can be reached using the technique of Appendix \ref{z2sec}.

 \para
The different phase also shows up in the spin-Hall conductivity. Using the TKNN formula  now gives $\sigma_{xy}=0$ for our model. 
Again, this result can also be derived directly, without reference to the underlying lattice,  by integrating out the fermions at finite chemical potential, this time in the presence of $B$ generating the Landau levels. 

\subsubsection*{Again, the View from the Boundary}

The essence of the $\mathbb{Z}_2$ topological insulator is that it exhibits an odd number of counter-propagating boundary modes when placed side-by-side with an ordinary insulator. We now demonstrate that this is indeed the case in our system: a varying chemical potential ensures that this is so.

\para
It is simplest to consider a chemical potential that varies along the $x$-direction, $\mu=\mu(x)$. We insist only that 
\be \mu(x)\rightarrow 0 \ \ \ \ {\rm as}\ x \rightarrow -\infty \nn \ee
so that the region on the left is in  the $\nu=1$ phase.

\para
Meanwhile,  we want to sit in the $\nu=0$ phase on the right. This is ensured if
 \be  \mu(x)>\max_{\vec p}\sqrt{ m^2 + \frac{1}{2}|W(2\vec{p}/B)|^2}\ \ \ \ {\rm as}\ x\rightarrow +\infty \nn \ee

\begin{figure}[!h]
  \begin{center}
    \includegraphics[width=0.9\textwidth]{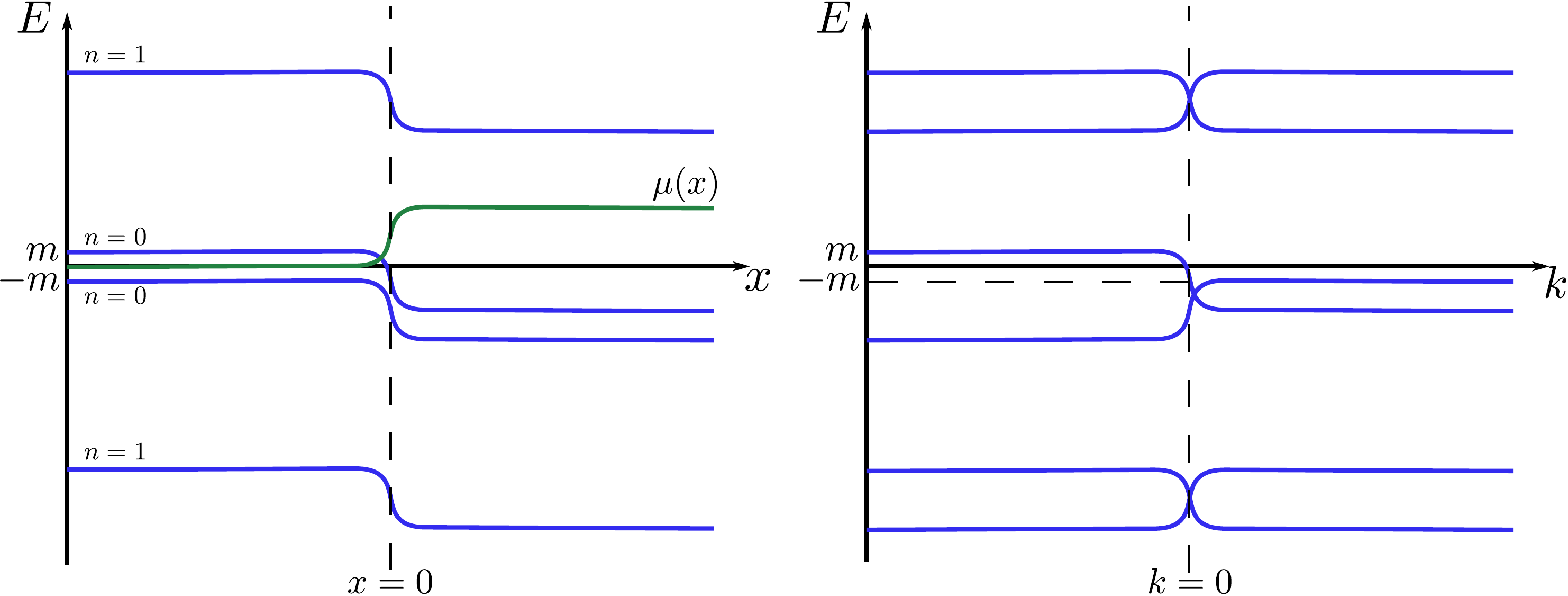}
  \end{center}
  \caption{Landau levels in the presence of a chemical potential step. The left panel shows the Landau levels and the chemical potential as a function of $x$. For a given $U(1)_A$ charge, each of the $n=1$ levels are twofold degenerate, while the $n=0$ levels are nondegenerate. The right panel shows the corresponding dispersion relation for a given $U(1)_A$ charge.}
  \label{edgefig}
\end{figure}

\para
To explain how the pair of counter-propagating modes arises, we first consider the system in the regime $B < m_W^2$ where the W-boson condensate is not present. (We will see shortly that the W-boson condensate cannot remove these boundary modes). The energy of the various Landau levels is shown in the left-hand panel of Figure \ref{edgefig}. 

\para
The fermion modes in Landau gauge given in  \eqn{lll1} and \eqn{lll2} provide a basis of Landau level operators, labelled by a real parameter $k$ that represents the momentum of the modes in the $y$ direction, along the domain wall. Each mode is narrowly centered in the $x$ direction around $x\sim \pm 2k/B$\footnote{A similar relation also applies to the higher Landau levels}. As long as the chemical potential varies on scales much larger than the width $\sim B^{-1/2}$ of the Laudau level modes, one can treat these Landau levels states as localised states, whose energies are offset by the local value of the chemical potential\footnote{In this context, it is important that we think of the chemical potential as a perturbation pushing the Landau levels with respect to a zero of energy common to the two sides of the domain wall, and not as a perturbation that introduces nonequilibrium populations on the two sides.}. This is depicted in the left panel of Figure~\ref{edgefig}. However, since  the position in the $x$-direction is correlated with the momentum in the $y$-direction through $x\sim  \pm 2k/B$, the $x$-dependence of the energies of the Landau levels can also be thought of as a dispersion relation in $y$ direction, parallel to the domain wall. This is depicted for one of the $U(1)_A$ sectors in the  right panel of Figure~\ref{edgefig}; the diagram for the complementary sector follows from relabelling $k\rightarrow -k$ in the Figure.  In the interpolation region, we have a pair of counter-propagating modes at zero energy: a left-mover from one $U(1)_A$ sector (seen in the Figure) and a right-mover from the other $U(1)_A$ sector (not shown in the Figure).

\para
Now let us take $B > m_W^2$ and reinstate the W-boson condensate. The boundary modes cannot be removed: for that to occur, one would need to either close the gap or couple the left- and right-movers. The W-bosons do neither. The pair of counter-propagating modes are there to stay, providing a different demonstration that  the two sides of the wall have different $\mathbb{Z}_2$ invariants.

\section{Summary}\label{sumsec}

It has been a long journey. In this final section, we present a brief precis of our  main results. 

\subsubsection*{Massless Fermions}

In the background of the W-boson lattice, massless fermions exhibit four Dirac points. The existence of these Dirac points is guaranteed by topological considerations \cite{volovikbook,wenzee,beri}. Perhaps more surprisingly, the location of these Dirac cones, at ``midway" points  in the Brillouin zone, is also dictated by symmetry. They lie at  magnetic Bloch momenta
\be (p_x,p_y) = \pm\left(\frac{\lambda \sqrt{B}}{8},\frac{\pi \sqrt{B}}{2\lambda}\right)\nn\ee
where $B$ is the background magnetic field and $\lambda$ is the parameter that determines the shape of the lattice.

\subsubsection*{Massive Fermions}

With the opening up of a gap, we can compute topological quantum numbers for different bands.  (These bands also carry $U(1)_F$ and $U(1)_A\subset SU(2)_F$ quantum numbers).

\para
Each individual band has a non-zero TKNN invariant $C_1 = \pm 1$. This result can be attributed to the same topological winding of the magnetic Bloch states that also gives rise to Dirac points in the  spectrum of massless fermions. However, the total TKNN invariant of the system vanishes, due to cancellations between contributions from bands related by time reversal symmetry.

\para
The total $\mathbb{Z}_2$ invariant of the system is non-vanishing, and this is manifest in a quantum spin-Hall effect in which a background electric field for $U(1)_F$ generates a transverse current for $U(1)_A$. However, the theory enjoys a charge conjugation symmetry $\tilde{\cal C}$ with $\tilde{\cal C}^2=-1$. This prohibits the existence of topologically distinct phases. Although our system enjoys a non-vanishing $\mathbb{Z}_2$ invariant, the same is true of all systems within this symmetry class. Our ``topological insulator" does not have a ``non-topological" partner with which to mate and produce massless edge states. 

\subsubsection*{Massive Fermions with Chemical Potential}

Adding a chemical potential for $U(1)_F$ breaks charge conjugation symmetry, freeing us from its prohibitive restrictions. By varying the chemical potential, it is simple to find phases with vanishing $\mathbb{Z}_2$ invariant to pair with our earlier state. Placing the two together results in massless edge states, protected by time reversal invariance, as expected.

\para
Finally, we finish with a comment on the importance of the $\mathbb{Z}_2$ invariant. For much of this paper, the computation of the $\mathbb{Z}_2$ invariant has not provided us any information beyond the TKNN invariants and spin-Hall conductivity. However, as stressed in many other contexts (see, e.g., the reviews \cite{hk,zreview}) the $\mathbb{Z}_2$ is primal: it is more robust than the Chern-Simons terms and spin-Hall conductivity. 

\para
To see this, suppose that we add to our system some interaction 
that breaks $U(1)_A$, while preserving $U(1)_F$ and the discrete symmetries and, moreover, keeps the gap open. In the absence of a conserved $U(1)_A$ current, there is now no meaning to the spin-Hall conductivity. Similarly, with no quantum numbers to distinguish different bands,  TKNN invariants can be defined only for time-reversed pairs of bands for which they vanish. Yet the $\mathbb{Z}_2$ invariants remain. 
Furthermore, the edge modes on the domain wall interpolating between different chemical potentials survive -- they are guaranteed to remain massless by Kramers' theorem. This phenomenology is not tied to the Chern-Simons terms, but owes its existence  to a more subtle topological order protected by time reversal symmetry. This is the essence of the $\mathbb{Z}_2$ topological insulator. It is pleasing that this simplest topological insulator sits naturally within the simplest $SU(2)$ gauge theory. 

\subsubsection*{Future Developments}

We have seen that the quantum spin-Hall effect and the 2d topological insulator sit naturally within $SU(2)$ Yang-Mills theories.  From the condensed matter perspective, there are a number of ways to construct non-dynamical pseudo-magnetic fields (see, for example, \cite{notashwin} for a method to using graphene). More generally, it is natural to ask whether the more subtle three-dimensional ${\bf Z}_2$ topological insulator has a natural home within non-Abelian gauge theories and how this may shed light on the phase structure of these theories.

\newpage

\section*{Acknowledgements}

We would like to thank Mike Blake, Michal Kwasigroch, Steve Simon, Andrei Starinets and Edward Witten for useful comments and discussions. BB acknowledges the support of a Marie Curie IEF Fellowship. DT and KW  are supported by STFC and by the European Research Council under the European Union's Seventh Framework Programme (FP7/2007-2013), ERC grant agreement STG 279943, ``Strongly Coupled Systems".

\appendix
\section{Appendix: Stability of the W-Boson Lattice}\label{trustapp}

The lattice solutions described in Section \ref{sec2} have been extensively studied in the literature and there exist both numerical constructions \cite{ao1andahalf} and analytic proofs of their existence  \cite{proof}. (These papers deal with lattices in electroweak theory but, in one limit, reduce to the model considered here and in \cite{ao0}). Here we would like to discuss some observations on the stability of these lattices\footnote{We thank Edward Witten for correspondence on this topic.}.

\para
The question at issue is whether there is some topological quantum number which can prevent the lattice from relaxing to the vacuum.
 In the case of Abelian gauge theories, a non-vanishing background $B$ field can be ensured in some region by insisting that $\oint A\cdot dl$ takes some fixed, non-zero value on a boundary. However, for non-Abelian gauge theories this argument does not work because the line integral is not gauge invariant. Moreover, in our Yang-Mills-Higgs model, the existence of monopole solutions -- which now play the role of instantons \cite{polyakov} -- means that the magnetic field inside a region can always relax non-perturbatively. It appears that the lattice is, at best, meta-stable.

\para
To our knowledge, the question of perturbative stability of the lattice has not been fully addressed. In closely related models, there are both Bogomolnyi type arguments \cite{ao1} and good numerical studies \cite{ao1andahalf}. However, these arguments always assume the existence of a discrete lattice symmetry and it is unclear if the solutions remain stable for perturbations outside of this ansatz. While it seems very plausible that the lattice is meta-stable, to our knowledge this remains an open question. However, for the purposes of the band-structure studies in this paper, we need only assume the existence of the lattice.

\section{Appendix: Identifying the Bloch Momenta}\label{blochsec}

Our goal in this section is to identify the magnetic Bloch momenta. This is achieved by understanding how translational symmetries act on the states of the system. In Section 2, we saw that the background gauge fields are invariant under translations by the magnetic unit cell only if accompanied by a gauge transformation. 

\para
Specifically, the magnetic spatial unit cell\footnote{Note that this differs from the lattice unit cell, generated by the vectors  $\vec{d}_i$ defined in Section \ref{wbosonsec}. The lattice unit cell  has half the area of the magnetic unit cell felt by the fermions.} compatible with the W-boson lattice is generated by the vectors $\vec{d}_1^{\,\prime} = (4\pi/\lambda\sqrt{B},0)$ and $\vec{d}_2^{\,\prime} = (0,\lambda/\sqrt{B})$.  The background fields are invariant under $\vec{x}\rightarrow \vec{x}+\vec{d}_2^{\,\prime}$. However, the transformation $\vec{x}\rightarrow \vec{x}+\vec{d}_1^{\,\prime}$ requires a compensating gauge transformation,
\be G = \exp \left ( - \frac{4\pi i \sqrt B  y}{\lambda} T^3 \right) \nn\ee
To determine the Bloch momenta of the states, we need to see how this combination of translation and gauge transformation acts on the creation and annihilation operators. After some algebra, we find under $\vec{x}\rightarrow \vec{x}+\vec{d}_2^{\,\prime}$, we have
\be a_i^\star(\vec{p}) \rightarrow e^{i\lambda p_1/\sqrt{B}}a_i^\star(\vec{p})\ \ \ \ {\rm and}\ \ \ \ b_i^\star(\vec{p}) \rightarrow e^{-i\lambda p_1/\sqrt{B}}b_i^\star(\vec{p})\nn\ee
Meanwhile, under the combination $\vec{x}\rightarrow \vec{x}+\vec{d}_1^{\,\prime}$, together with the gauge transformation $G$, we have
\be a_i^\star(\vec{p}) \rightarrow e^{4\pi ip_2/\lambda\sqrt{B}}a_i^\star(\vec{p})\ \ \ \ {\rm and}\ \ \ \ b_i^\star(\vec{p}) \rightarrow e^{-4\pi ip_2/\lambda\sqrt{B}}b_i^\star(\vec{p})\nn\ee
We conclude that the states $a^\star_i(p_1,p_2) \vert \Omega \rangle$ have Bloch momentum $(p_x,p_y)=(p_2,p_1)$. Note the interchange of indices! Meanwhile, the states $b^\star_i(p_1, p_2) \vert \Omega \rangle$ have Bloch momentum $(p_x,p_y) = -(p_2,p_1)$

\subsubsection*{An Ambiguity}

Note that we could have chosen to accompany the translation in the  $\vec{d}_1^{\,\prime}$ direction by the gauge transformation, 
\be \tilde{G} = \exp \left ( - \frac{2\pi i \sqrt B  y}{\lambda} T^3 + 2\pi i T^3\right) \nn\ee
This too leaves the W-boson lattice invariant, but has the effect of putting a minus sign in front of the spinors. This has the effect of changing the Bloch momenta of the states. The ambiguity can be thought of as a discrete choice of origin for Bloch momenta.  The upshot is that the Bloch momenta only have invariant meaning modulo shifts by $(p_x,p_y)=(\Delta p_2 / 2,0)$ and $(p_x,p_y) = (0,\Delta p_1 / 2)$.

\section{Appendix: Diagonalising the Fermions}\label{diagonalapp}

In this appendix, we explain how to diagonalise the fermionic Hamiltonian in the presence of the lattice. Specifically, our goal is to show that the diagonal Hamiltonian \eqn{diagonal}
\be H_{\rm diag} = \sum_{i=1}^2\int_{\mathbf T^2} d^2p \ \left[ iW\left(\frac{2p_1}{B},\frac{2p_2}{B}\right)\, \zeta_i^{-}(\vec{p})^\star\,\zeta^+_i(\vec{p}) + {\rm h.c.} \right]\label{diagonal2}\ee
is equivalent to the Hamiltonian \eqn{deltah} that arises from placing the lowest Landau level fermions in the background of the W-boson lattice
\be
H_{\rm lattice}  =   \sum_{i=1}^2 \int \frac{dk}{2\pi} \frac{dk'}{2\pi}  \left[ i \phi(k,k')\ \xi^{-}_i(k')^\star\,\xi_i^+(k) + {\rm h.c.}\right]\label{deltah1}\ee
where $\phi(k,k')$ is given by
\be \phi(k,k') &=&\sqrt{\frac{B}{2\pi}} \int dxdy\   \exp \left( i(k'+k)y - \frac B 4 \left( x + \frac {2k} B \right)^2 - \frac B 4 \left( x + \frac {2k'} B \right)^2 \right)W(x,y)\nn\ee
The Grassmann operators $\zeta$ are related to  $\xi$ by the linear map
\be
\zeta^+_i(p_1,p_2) &=& \frac 1 {\sqrt{2\pi \Delta p_2}} \sum_{n = -\infty}^\infty e^{2\pi i n p_2/\Delta p_2} \xi^+_i
(+p_1 + n\Delta p_1)\nn\\
\zeta^-_i(p_1,p_2) &=& \frac 1 {\sqrt{2\pi \Delta p_2}} \sum_{n = -\infty}^\infty e^{2\pi i n p_2/\Delta p_2}\, \xi^-_i(-p_1+ n\Delta p_1)
\nn \ee
To demonstrate this, we start by substituting these expressions for $\zeta$, together with the linearised W-boson lattice \eqn{abrikosov} into  \eqn{diagonal2}. To simplify the expressions, we will drop the $i=1,2$ index and the ``${\rm h.c.}$", both of which simply go along for the ride. We have
\be H_{\rm diag} = \frac{i \bar W}{2\pi \sqrt 2 \Delta p_2}\int_{{\bf T}^2}d^2p \sum_{m,n,n'} &&e^{-i\pi m^2/2}e^{2\pi i p_2 (n-n'-m)/\Delta p_2}e^{- \frac 1 {2B} (2p_1 + m\Delta p_1)^2}
\nn\\ &&\ \times\,\xi^-(-p_1+n'\Delta p_1)^\star\,\xi^+(p_1+n\Delta p_1)\nn\ee
At this stage, we perform the integral over $p_2$. It  results in a Kronecker delta, imposing $m=n-n'$. In the following, we leave $m$ as a variable but the equality $m=n-n'$ is assumed. We find
\be H_{\rm diag} = \frac{i \bar W}{2\pi \sqrt 2 }\int_{-\Delta p_1/2}^{+\Delta p_1/2}dp_1 \sum_{n,n'} &&e^{-i\pi m^2/2}e^{- \frac 1 {2B} (2p_1 + m\Delta p_1)^2}
\nn\\ &&\ \times\,\xi^-(-p_1+n'\Delta p_1)^\star\,\xi^+(p_1+n\Delta p_1)\nn\ee
We now interchange the order of the integral and sum and define the new variable integration variable $k' = -p_1 - m\Delta p_1/2$. The Hamiltonian becomes
\be  H_{\rm diag} = \frac{i\bar W} {\sqrt 2} \sum_{n,n'} \int_{I_m} \frac{dk'}{2\pi}  &&e^{-i\pi m^2/2}e^{-2{k'}^2/B} 
\nn\\ &&\ \times\,\xi^-(k'+\Delta p_1(n+n')/2)^\star\,\xi^+(-k'+\Delta p_1(n+n')/2)\nn\ee
where the integral is taken over the interval 
\be I_m =( \Delta p_1/2)[-m-1,-m+1]\nn\ee
%
%
%
We now define a new summation parameter $q=n+n'$. We would like to convert the sum over $n$ and $n'$ into a sum over $q$ and $m=n-n'$. We should be a bit careful since if $q$ is odd then $m$ is necessarily odd and vice versa. This observation also allows us to replace $e^{-i\pi m^2/2}$ (which is either $1$ or $-i$) with $e^{-i\pi q^2/2}$. We have
\be H_{\rm diag} = \frac{i\bar{W}}{\sqrt 2}\left(\sum_{q,m\ {\rm odd}} + \sum_{q,m\ {\rm even}}\right) && \int_{I_m} \frac{dk'}{2\pi}  e^{-i\pi q^2/2}e^{-2k^{\prime\,2}/B} 
\nn\\ &&\ \times\,\xi^-(k'+q\Delta p_1/2 )^\star\,\xi^+(-k'+q\Delta p_1/2)\nn\ee
But now each of the sums over $m$ is trivial; they simply extend the range of the $k'$ integral,
\be \sum_{m\ {\rm odd}}\int_{I_m} dk' = \sum_{m\ \rm{even}}\int_{I_m}dk' = \int_{-\infty}^{+\infty}dk'\nn\ee
and we have the happy expression,
\be  H_{\rm diag} = \frac{i\bar{W}}{\sqrt 2}\sum_q \int \frac{dk'}{2\pi}  e^{-i\pi q^2/2}e^{-2k^{\prime\,2}/B} \,\xi^-(k'+q\Delta p_1/2)^\star\,\xi^+(-k'+ q\Delta p_1/2)\nn\ee
Now we're done with the hard work. It remains to show that $H_{\rm lattice}$, defined in \eqn{deltah1}, also takes this form. This requires nothing more complicated than Gaussian integrals. Substituting the various quantities into \eqn{deltah1}, performing the $\int dy$ gives rise to a delta function which we use to kill $\int dk$ and $\int dx$ is simply a Gaussian. After performing these steps, followed by an obvious substitution, we find
\be H_{\rm diag} = H_{\rm lattice} \nn\ee
as promised.

\section{Appendix: Charge Conjugation and Time Reversal}\label{dissec}

In this appendix, we describe in some detail the discrete symmetries of the fermions propagating in the background of the W-boson lattice. This means that we would like to construct discrete symmetries of the theory which leave the magnetic field $B$, the W-boson lattice $W$ and the Higgs field  $\phi$ invariant.

\subsubsection*{Charge Conjugation}

With the choice of gamma matrix basis, $\gamma^\mu = (i\sigma^3,\sigma^1,\sigma^2)$, charge conjugation in $d=2+1$ is taken to act as
\be {\cal C}: \psi \rightarrow \gamma^1\psi^\star\ \ \ \ , \ \ \ \ {\cal C}: A_\mu \rightarrow - A_\mu^\star \ \ \ \ , \ \ \ \ {\cal C}: \phi \rightarrow - \phi^\star \nn\ee
However, the transformation on $A_\mu$ leaves neither the magnetic field nor the W-boson lattice invariant. For this reason, we accompany charge conjugation with the gauge transformation $G_2=e^{i\pi T^2}$ and define the new charge conjugation operator $\tilde{\cal C} = G_2\circ {\cal C}$. This combination leaves the gauge field and Higgs field untouched, and acts only on the fermions,
\be \tilde{\cal C}: \psi \rightarrow G_2\gamma^1\psi^\star \label{tildec}\ee
On the $n=0$ Landau level creation operators $\tilde{\cal C}$ acts as
\be \tilde{\cal C}:&& a^\star_1(\vec{p})\rightarrow +b_1^\star(-\vec{p}) \ \ ,\ \ b_1^\star(\vec{p})\rightarrow -a_1^\star(-\vec{p})\nn\\
&& a^\star_2(\vec{p})\rightarrow -b_2^\star(-\vec{p}) \ \ ,\ \ b_2^\star(\vec{p})\rightarrow +a_2^\star(-\vec{p})
\nn\ee
Note that we have  $\tilde{\cal C}^2=-1$, a relation that will be important later on. The fact that we are forced to define charge conjugation that squares to minus one can be traced to presence of the W-boson condensate, rather than the background magnetic field.

\subsubsection*{Time Reversal}

Time reversal ${\cal T}$ is an antiunitary operator: ${\cal T}(i){\cal T}^{-1} = -i$. Of course, it acts on spacetime as $(t,x,y)\rightarrow (-t,x,y)$ and this change will be implicit in the formulae below. Since the fermion bilinear $\bar{\psi}\psi$ is odd under the usual time reversal symmetry in $d=2+1$ dimensions, the presence of the fermion mass term \eqn{smass} means that we can define a good time reversal symmetry only if it is accompanied by an exchange of the fermion species. It is simple to check that the theory is invariant under
\be {\cal T}: &&\psi_1\rightarrow \gamma^2\psi_2\ \ \ ,\ \ \ \psi_2\rightarrow \gamma^2\psi_1 \nn\\
{\cal T}: && A_t\rightarrow A_t\ \ \ \ ,\ \ \ \ A_i\rightarrow -A_i \ \ \ ,\ \ \  \phi \to - \phi \nn\ee
However, once again we require a symmetry which acts only on the fermions, leaving the gauge and Higgs fields invariant. This can again be achieved by a compensating gauge transformation, $\tilde{\cal T}=G_2\circ{\cal T}$. The result is that the Higgs field and the gauge components $A_i^a$ are left invariant while the fermions transform as
\be  \tilde{\cal T}: &&\psi_1\rightarrow G_2\gamma^2\psi_2\ \ \ ,\ \ \ \psi_2\rightarrow G_2\gamma^2\psi_1 \nn\ee
%
On the fermion creation operators, time reversal acts as 
\be \tilde{\cal T}: && a^\star_1(\vec{p}) \rightarrow +ia^\star_2(-\vec{p})\ \ \ ,\ \ \ b^\star_1(\vec{p})\rightarrow -ib_2^\star(-\vec{p})\nn\\
&& a^\star_2(\vec{p}) \rightarrow +ia^\star_1(-\vec{p})\ \ \ ,\ \ \ b^\star_2(\vec{p})\rightarrow -ib_1^\star(-\vec{p})\label{t}\ee
With this definition, we have $\tilde{\cal T}^2=+1$. (In verifying this, it's crucial to remember that ${\cal T}$ is antiunitary.) However, in contrast to the choice of charge conjugation, there is an ambiguity in our choice of time reversal. We could define a new time reversal symmetry which mixes $\tilde{\cal T}$ with a discrete ${\bf Z}_2\in U(1)_A\times U(1)_F$ which acts as
%
\be {\bf Z}_2: \psi_1\rightarrow \psi_1 \ \ \ {\rm and}\ \ \ \psi_2\rightarrow -\psi_2\nn\ee
%
We can then define an alternative time reversal symmetry 
$\tilde{\cal T}^{\,\prime} = Z_2\circ \tilde{\cal T}$.
It is simple to check that this obeys $\tilde{\cal T}^{\,\prime\,2}=-1$.
Acting on the fermion  creation operators, we have
\be \tilde{\cal T}^\prime: && a^\star_1(\vec{p}) \rightarrow +ia^\star_2(-\vec{p})\ \ \ ,\ \ \ b^\star_1(\vec{p})\rightarrow -ib_2^\star(-\vec{p})\nn\\
&& a^\star_2(\vec{p}) \rightarrow -ia^\star_1(-\vec{p})\ \ \ ,\ \ \ b^\star_2(\vec{p})\rightarrow +ib_1^\star(-\vec{p})\label{tprime}\ee
Such ambiguity in the presence of other global $U(1)$ symmetries is not uncommon. 

\subsubsection*{Symmetry Translation Service} \label{subtlesec}

There is a seeming discrepancy between the language used in high-energy physics and the language used in condensed matter. The issue is the definition of charge conjugation. As described, our system enjoys a charge conjugation symmetry ${\cal C}$. As usual in high-energy theory, this is  a  unitary operation. Yet the classification of topological insulators \cite{altland}-\cite{ryu} is in terms of an {\it anti-unitary} charge conjugation operator. What's going on?

\para
The key is to distinguish between Fock space and first quantised operators \cite{ryu}. 
In high-energy physics, the action of operators such as time reversal ${\cal T}$ and charge conjugation ${\cal C}$ are always implicitly stated in terms of the Fock space. In contrast, 
in condensed matter,  theories of free fermions  are classified according  to the anti-unitary symmetries of the first quantised Hamiltonian, that is the Hermitian kernel $h(x-y)$ of the fermion action
%
\be
S=\int dt d^2x\ i\psi^\dagger \partial_t \psi - \int dt d^2x d^2y\ \psi^\dagger(x) h(x-y) \psi(y)\nn
\ee
Time reversal is anti-unitary on the Fock space, and also translates to an anti-unitary symmetry of $h(x-y)$ with the same square. However, charge conjugation is unitary on the Fock space. But, in terms of $h(x-y)$, it translates to the anti-unitary relation 
\be \hat{\cal C}\, h(x-y) \hat{\cal C}^{-1}=-h(x-y)\nn\ee
For our particular theory, $\hat{\cal C}$ descends from the action $\tilde{\cal C}$ defined in \eqn{tildec}, 
\be \hat{\cal C} = G_2\circ \gamma^1\circ K\nn\ee
where $K$ is complex conjugation. 
In terms of the classification table,  $\hat{\cal C}$ is referred to as particle-hole symmetry (much as it is in particle physics). 
Moreover, and most importantly, $\hat{\cal C}^2 = \tilde{\cal C}^2 =-1$.


\section{Appendix: $\mathbb{Z}_2$ Invariant from Inversion Symmetry}\label{z2sec}

Much of the recent excitement about topological insulators can be traced to the understanding that there are more subtle topological invariants beyond the Chern numbers. In time reversal invariant $2+1$ dimensional insulators, the appropriate quantity is a $\mathbb{Z}_2$ invariant \cite{km,fukanepump,moorebalents,fk}. 
%
In general, this
can be fairly tricky to compute. 
However, when the lattice enjoys an inversion symmetry ${\cal I}$, there is a particularly simple formula due to Fu and Kane \cite{fk}. Here we first demonstrate the existence of the inversion symmetry and subsequently detail the procedure for computing the $\mathbb{Z}_2$ invariant.

\para
Under a simple inversion, ${\cal I}:(x,y)\rightarrow (-x,-y)$, the W-boson lattice is left invariant: $W(\vec{x})=W(-\vec{x})$. However, implementing such an action on the fermions needs a little care. The equations of motion are invariant under the inversion symmetry ${\cal I}$ whose action on the fields is given by
\be {\cal I}: \psi_i \rightarrow \gamma^0\psi_i\ \ \ {\rm and}\ \ \ {\cal I}: A_t\rightarrow  A_t \ \ , \  \ A_i\rightarrow -A_i\nn\ee
To compensate for the change in the gauge field, we need to accompany this by a gauge transformation $G_3 = e^{-i\pi T^3}$. The resulting inversion symmetry $\tilde{\cal I} = G_3\circ {\cal I}$ leaves the background gauge potential unchanged but acts on the fermions. $\tilde{\cal I}$ flips the momenta of the fermion states,
\be \tilde{\cal I}:&& a^\star_i(\vec{p})\rightarrow a_i^\star(-\vec{p}) \ \ ,\ \ b_i^\star(\vec{p})\rightarrow b_i^\star(-\vec{p})\nn
\nn\ee

\para
We now move on to calculate the $\mathbb{Z}_2$ invariant following the Fu-Kane prescription. We pick one band from the filled pair of  time reversal conjugates. We choose to work with the   $b_1|0\rangle$ band. 
Next, find the points in the Brillouin zone that are preserved by the inversion symmetry. These are $\vec{p} = (0,0)$, $\vec{p}=  (\frac 1 2 \Delta p_1, 0)$, $\vec{p}=(0, \frac 1 2 \Delta p_2 )$ and $\vec{p}=(\frac 1 2 \Delta p_1, \frac 1 2 \Delta p_2)$.

\para
Now, find the eigenvalues of the states at these four points under inversion. These eigenvalues will always be $\pm 1$. Using the transformation properties of $b_1(\vec{p})$ 
 under lattice translations \eqn{abtrans}, we have
 \be \tilde{\cal I}&:& b_1(0,0)\ \rightarrow\  b_1(0,0)\nn\\
 \tilde{\cal I}&:& b_1(\Delta p_1 / 2, 0) \  \rightarrow\  b_1(\Delta p_1 / 2, 0)\nn\\
 \tilde{\cal I}&:& b_1(0, \Delta p_2 / 2)\ \rightarrow  \ b_1(0, \Delta p_2 / 2)\nn\\
 \tilde{\cal I}&:& b_1(\Delta p_1 / 2, \Delta p_2 / 2)\  \rightarrow \ - b_1(\Delta p_1 / 2, \Delta p_2 / 2)\nn\\
  \nn\ee
 The $\mathbb{Z}_2$ invariant -- usually denoted $(-1)^\nu$ -- is the product of these eigenvalues. For the band $b_1|0\rangle$, this invariant  is given by
\be
(-1)^\nu =  1\times 1\times 1\times (-1) = -1
\nn\ee
We thus find $\nu = 1$, in accordance with the result of the main text.

\end{document}